\documentclass[]{pasj02} 
\usepackage[switch,mathlines]{lineno} 
\usepackage{xcolor}

\jyear{2024}
\Received{}
\Accepted{}

\begin{document} 

\title{XMM-Newton observations of ten high-redshift CAMIRA clusters of galaxies}

\author{
Naomi \textsc{Ota},\altaffilmark{1,2}\altemailmark \orcid{0000-0002-2784-3652} \email{naomi@cc.nara-wu.ac.jp} 
Ikuyuki \textsc{Mitsuishi},\altaffilmark{3} \orcid{0000-0002-9901-233X}
Nobuhiro \textsc{Okabe},\altaffilmark{4,5,6} \orcid{0000-0003-2898-0728}
Masamune \textsc{Oguri},\altaffilmark{7,8} \orcid{0000-0003-3484-399X}
Yoshiki \textsc{Toba},\altaffilmark{9,1,11,12}\orcid{0000-0002-3531-7863}
Kuga \textsc{Harada},\altaffilmark{3}
Marie \textsc{Kondo},\altaffilmark{13}\orcid{0009-0005-5685-1562}
Satoshi \textsc{Miyazaki},\altaffilmark{10} \orcid{0000-0002-1962-904X}
Koki \textsc{Sakuta},\altaffilmark{3}\orcid{0009-0005-2657-2554}
Kosuke \textsc{Sato},\altaffilmark{13,14}\orcid{0000-0001-5774-1633}
Anri \textsc{Yanagawa},\altaffilmark{1}\orcid{0009-0009-3388-2509} 
Anje \textsc{Yoshimoto}\altaffilmark{1}
}
\altaffiltext{1}{Department of Physics, Nara Women’s University, Kitauoyanishi-machi, Nara, Nara 630-8506, Japan}
\altaffiltext{2}{Argelander-Institut f\"{u}r Astronomie (AIfA), Universit\"{a}t Bonn, Auf dem H\"{u}gel 71, 53121 Bonn, Germany}
\altaffiltext{3}{Department of Physics, Nagoya University, Aichi 464-8602, Japan}
\altaffiltext{4}{Department of Physical Science, Hiroshima University, 1-3-1 Kagamiyama, Higashi-Hiroshima, Hiroshima 739-8526, Japan}
\altaffiltext{5}{Hiroshima Astrophysical Science Center, Hiroshima University, 1-3-1, Kagamiyama, Higashi-Hiroshima, Hiroshima 739-8526, Japan}
\altaffiltext{6}{Core Research for Energetic Universe, Hiroshima University, 1-3-1, Kagamiyama, Higashi-Hiroshima, Hiroshima 739-8526, Japan}
\altaffiltext{7}{Center for Frontier Science, Chiba University, 1-33 Yayoi-cho, Inage-ku, Chiba 263-8522, Japan}
\altaffiltext{8}{Department of Physics, Graduate School of Science, Chiba University, 1-33 Yayoi-Cho, Inage-Ku, Chiba 263-8522, Japan}
\altaffiltext{9}{Department of Physical Sciences, Ritsumeikan University, 1-1-1 Noji-higashi, Kusatsu, Shiga 525-8577, Japan}
\altaffiltext{10}{National Astronomical Observatory of Japan, 2-21-1 Osawa, Mitaka, Tokyo 181-8588, Japan}
\altaffiltext{11}{Academia Sinica Institute of Astronomy and Astrophysics, 11F of Astronomy-Mathematics Building, AS/NTU, No.1, Section 4, Roosevelt Road, Taipei 10617, Taiwan}
\altaffiltext{12}{Research Center for Space and Cosmic Evolution, Ehime University, 2-5 Bunkyo-cho, Matsuyama, Ehime 790-8577, Japan}
\altaffiltext{13}{Department of Physics, Saitama University, 255 Shimo-Okubo, Sakura-ku, Saitama, Saitama 338-8570, Japan}
\altaffiltext{14}{Department of Astrophysics and Atmospheric Sciences,
Kyoto Sangyo University, Kamigamo-motoyama, Kita-ku, Kyoto 603-8555, Japan}



\KeyWords{cosmology: observations --- galaxies: clusters: intracluster medium --- X-rays: galaxies: clusters --- galaxies: active}

\maketitle

\begin{abstract}
We present results from XMM-Newton observations of ten high-redshift ($0.81 < z < 1.17$) galaxy clusters selected from the CAMIRA catalog based on high richness ($N > 40$). These massive clusters, identified in the Hyper Suprime-Cam Subaru Strategic Program field, provide an ideal sample for probing the dynamical state of the intracluster medium (ICM) in the early Universe. We performed uniform X-ray imaging and spectral analyses to measure the ICM temperature and bolometric luminosity, and investigated cluster morphology through offsets between the brightest cluster galaxy (BCG) and the X-ray peak. Extended X-ray emission was detected from all targets, but only one system was classified as dynamically relaxed, indicating a low relaxed fraction ($\sim 10\%$) at high redshift.

By combining this high-$z$ sample with a lower-redshift CAMIRA cluster sample, we derived scaling relations among richness, temperature, luminosity, and mass. The results are broadly consistent with predictions from both the self-similar model and the baseline model incorporating the mass--concentration relation. We find no significant redshift evolution, strengthening the view that cluster scaling relations are largely established by $z \sim 1$. We also examined the AGN fraction among member galaxies and found significantly higher AGN activity in high-redshift clusters, particularly in the outskirts, suggesting enhanced AGN triggering during early cluster assembly and a possible connection to the thermodynamic state of dynamically young clusters. These findings provide new insights into the formation and evolution of massive clusters and the thermodynamic history of the ICM, and complement large-area X-ray surveys such as eROSITA.
\end{abstract}


\section{Introduction}\label{sec:intro}
Galaxy clusters, the most massive gravitationally bound systems in the Universe, serve as powerful cosmological probes and laboratories for studying galaxy evolution and the thermal history of the intracluster medium (ICM). In particular, observations of high-redshift clusters offer a unique window into early structure formation and the evolution of ICM properties.

X-ray observations provide not only constraints on the thermodynamic properties of the intracluster medium, but also key insights into the dynamical states of galaxy clusters. In particular, structural indicators such as the offset between the brightest cluster galaxy (BCG) and the X-ray peak serve as valuable diagnostics of cluster assembly and merger history \citep{Katayama03, Sanders09}.

Previous X-ray studies of high-redshift clusters have shown that large BCG--X-ray offsets and disturbed morphologies are common at $z \gtrsim 1$, suggesting that many systems are dynamically young (e.g., \cite{Fassbender11}). On the other hand, Chandra observations of extremely massive clusters selected by the Sunyaev-Zel’dovich (SZ) effect indicate that the average X-ray morphology and ICM structure exhibit little evolution out to $z \sim 1.9$ \citep{McDonald17}. These apparently contrasting results highlight the importance of cluster mass and selection effects when interpreting morphological indicators at high redshift.

Complementary to studies of cluster dynamics, the redshift evolution of X-ray scaling relations provides important constraints on the thermodynamic assembly of the ICM. X-ray measurements of the ICM temperature and luminosity enable the study of scaling relations with mass, which have been widely used in cluster cosmology (e.g., \cite{Pratt19, Giodini13}). More recently, weak-lensing (WL) measurements have become a primary approach for mass calibration in cosmological studies (e.g., \cite{Chiu22, Bocquet24a,Bocquet24b,Grandis24, Kleinebreil25,Ghirardini24}), offering an independent and complementary method to X-ray analyses. Previous studies have reported mixed results on the redshift evolution of these relations. \citet{Reichert11} found only weak evolution, although their heterogeneous cluster sample may be affected by selection biases. In contrast, \citet{Bulbul19} reported significant deviations from the self-similar $L-M$ relation in SZ-selected clusters, with evolution that remains poorly constrained. More recently, \citet{Ota23} analyzed optically selected CAMIRA clusters located in the eROSITA Final Equatorial-Depth Survey (eFEDS) field and found consistency with both the self-similar model and the revised baseline model of \citet{Fujita19}, while a stacking analysis of lower-richness CAMIRA systems \citep{Nguyen25} provides further support. These studies jointly suggest that although selection effects may influence the observed scaling behaviors, the thermodynamic structure of massive clusters is broadly established by $z \sim 1$, with possible evolution in normalization or scatter.

Simulations incorporating active galactic nucleus (AGN) feedback predict early ICM heating and morphological disturbances \citep{Muanwong06, Truong18}. Observationally, the AGN fraction among cluster galaxies increases with redshift \citep{Toba24,Hashiguchi23}, implying that AGN activity is more prevalent in dynamically young systems at high $z$.

The Subaru Hyper Suprime-Cam Strategic Program (HSC-SSP; \cite{HSC1styr, HSC1styrOverview}) has enabled the identification of thousands of clusters using the CAMIRA algorithm \citep{Oguri14b, Oguri18}. At the same time, the eROSITA all-sky survey has discovered a large number of high-$z$ clusters \citep{Bulbul24}, but its shallow depth often limits the photon statistics required for reliable temperature measurements. Deep, targeted X-ray observations are therefore essential to robustly characterize the ICM properties of high-redshift clusters.

Motivated by this, we conducted XMM-Newton observations of ten high-richness ($\hat{N}_{\mathrm{mem}} > 40$) CAMIRA clusters at $z > 0.8$ from the S16A catalog. We perform uniform X-ray imaging and spectral analyses and combine these clusters with a lower-redshift CAMIRA sample from \citet{Ota20}. The goals of this study are threefold:

\begin{enumerate}
\item to characterize the dynamical states of the clusters via BCG-X-ray offset measurements (section~\ref{subsec:peakoffset});
\item to investigate scaling relations among richness, temperature, luminosity, and mass, and to assess their redshift evolution across $0.14 < z < 1.17$ (section~\ref{subsec:scaling});
\item to evaluate the AGN fractions of member galaxies and examine their dependence on redshift and cluster-centric radius (section~\ref{subsec:agn_fraction}).
\end{enumerate}

By focusing on a uniformly selected high-$z$ CAMIRA sample with deep XMM-Newton data, this study provides complementary insights to wide-area surveys such as eROSITA and offers a detailed view of the physical state and evolution of massive clusters in the early Universe. While we assess the prevalence of AGN activity, its direct thermal impact on the ICM is beyond the scope of this work.

Throughout this paper, we adopt a $\Lambda$CDM cosmology with $\Omega_{\rm M}=0.28$, $\Omega_\Lambda=0.72$, and $h=0.7$. Elemental abundances follow \citet{Lodders09}, and all uncertainties are quoted at the $1\sigma$ level unless otherwise noted.

\section{Sample}
\begin{figure*}[htb]
 \begin{center}
 \includegraphics[width=0.45\linewidth,angle=0]{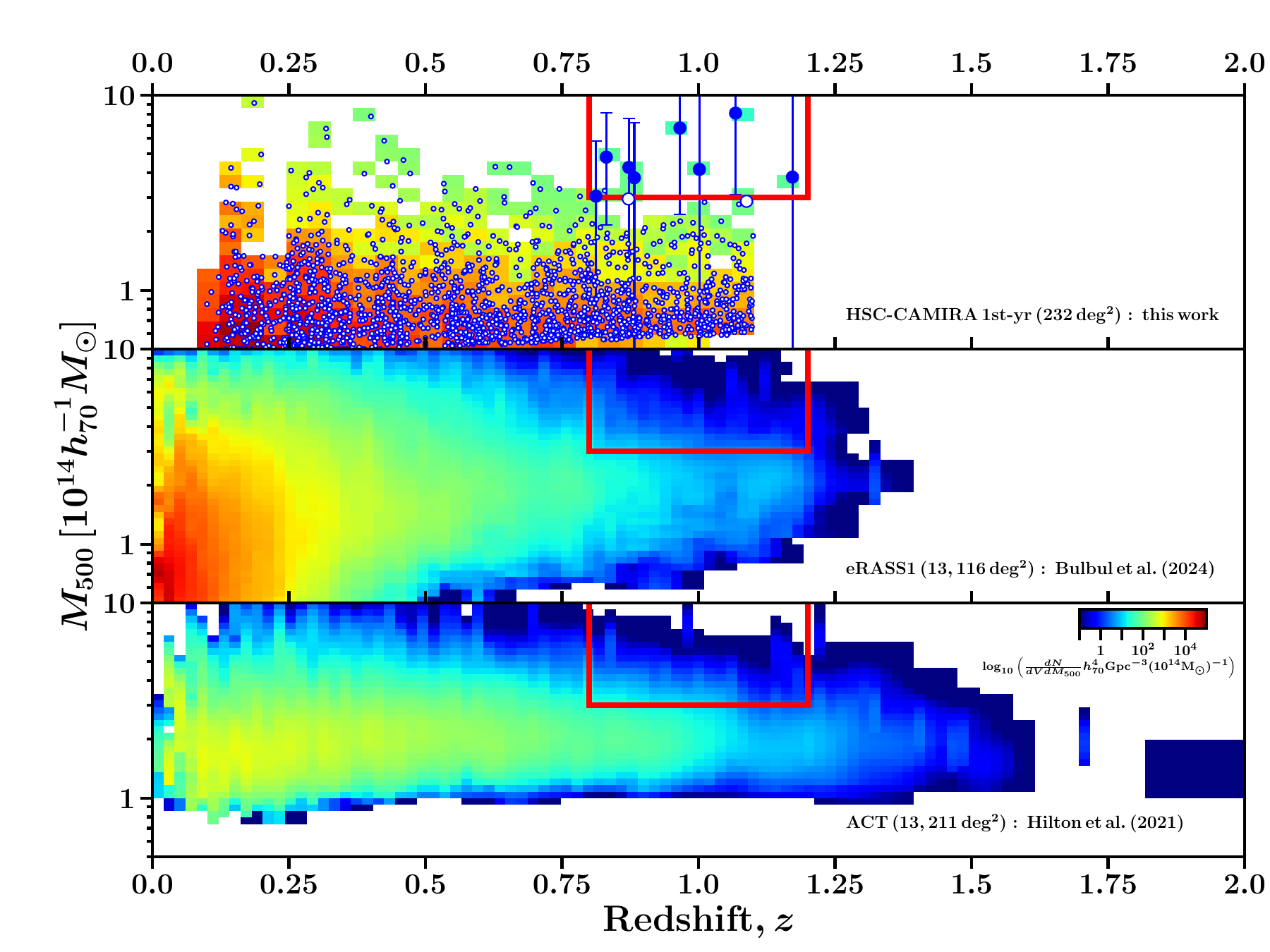}
 \includegraphics[width=0.45\linewidth,angle=0]{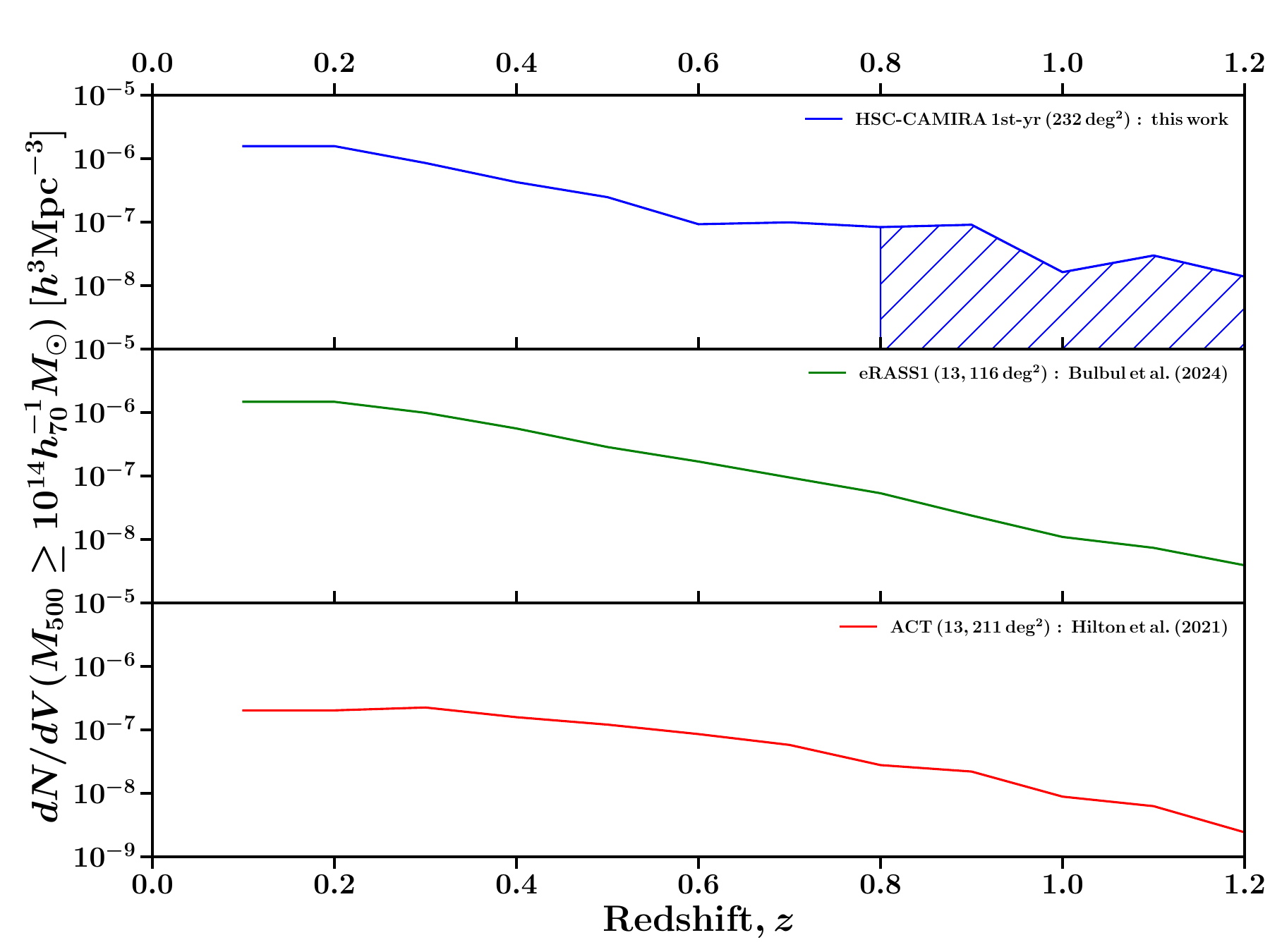}
 \end{center}
 \caption{{\it Left}: Mass versus redshift for the optically-selected clusters (Top; \cite{Oguri18}), X-ray-selected clusters (middle, eRASS1; \cite{Bulbul24}), and SZ-selected clusters (bottom, ACT; \cite{2021ApJS..253....3H}). The background colors represent $\log_{10}\left(dN/(dVdM)\,h_{70}^4{\rm Gpc}^{-3}(10^{14}M_\odot)^{-1}\right)$. The red solid boxes correspond to the selection function of the present sample. The blue open and filled circles denote the masses inferred by cluster richness and WL mass, respectively. {\it Right}: Number density of clusters per a co-moving volume of 1 $h^{-3}{\rm Mpc}^3$, where the mass uncertainties are not taken into account. The hatched region indicates the redshift range covered by the cluster sample analyzed in this work.
 {Alt text: Two-panel figure. The left panel shows three scatter plots of cluster mass versus redshift for optically selected, X ray selected, and Sunyaev Zeldovich selected samples, overlaid on a color coded density map. Red boxes mark the selection region of the present sample, and circles indicate richness based and weak lensing based mass estimates. The right panel shows a line graph of cluster number density as a function of mass and redshift. }
}\label{fig:n-z} 
\end{figure*}

Figure~\ref{fig:n-z} shows the redshift distribution of galaxy clusters identified by the CAMIRA algorithm in the HSC-SSP survey (hereafter, the CAMIRA clusters; \cite{Oguri18}). The number density indicates that the HSC survey is highly effective at detecting massive clusters even at high redshift. In parallel, the recently released first eROSITA all-sky survey \citep{Bulbul24} has discovered more than an order of magnitude more high-$z$ clusters. Within $0.8 < z < 1.2$, 157 X-ray-selected clusters with $M_{500} > 3 \times 10^{14}\, h^{-1} M_{\odot}$ have been identified, although reliable temperature measurements are available for only 42 of them. Because most eROSITA detections contain fewer than $\sim 100$ photons, their temperature estimates suffer from large statistical uncertainties. Deep X-ray observations are therefore essential to obtaining accurate ICM thermodynamic properties for high-redshift clusters and to complement the eROSITA survey.

Given this context, we focused on clusters with richness $\hat{N}_{\mathrm{mem}}$ greater than 40 (or a mass greater than $3\times10^{14}~h^{-1} M_{\odot}$) and a redshift greater than 0.8. Throughout this paper, we denote the CAMIRA catalog richness as $\hat{N}_{\mathrm{mem}}$ and use a simplified notation $N$ when discussing scaling relations. From 1,921 objects in the CAMIRA S16A catalog \citep{Oguri18}, we identify ten clusters that satisfy these criteria (table~\ref{tbl:sample}) and obtained XMM-Newton observations with exposure times of 16--60~ks. The net exposure times after data screening (section~\ref{subsec:reduction}) are listed in table~\ref{tbl:spec}.

To study redshift evolution of cluster properties, we additionally use a low-redshift comparison sample consisting of 17 CAMIRA clusters with high-quality X-ray data from \citet{Ota20}. These clusters span $0.14 < z < 0.75$ with richness values of $20 < \hat{N}_{\mathrm{mem} }< 70$. The combination of the high-$z$ and low-$z$ samples allows us to investigate cluster scaling relations and AGN activity over a broad redshift range of $0.14 < z < 1.17$.

\begin{table*}[htb]
  \tbl{Sample list.}{%
  \begin{tabular}{lllllllll}\hline\hline
Cluster  & $z$ & $\hat{N}_{\rm mem}$$^{\mathrm{a}}$ & $R_{500}$&  BCG position & X-ray centroid & X-ray peak & $D_{\rm XC}$$^{\mathrm{b}}$ & $D_{\rm XP}$$^{\mathrm{c}}$  \\ 
	    & & & (Mpc/\arcsec) & RA, Dec (deg)  & RA, Dec (deg) &RA, Dec (deg) & (kpc) & (kpc)  \\ \hline
HSCJ115653-003807 & 0.812 & 64.7 & 0.871/114 & 179.2208,-0.6354 & 179.2248,-0.6335 & 179.2248,-0.6327 & $123\,(\pm 38)$ & 132 (82, 208) \\
HSCJ144933-004301 & 0.966 & 52.0 & 0.749/93 & 222.3887,-0.7170 & 222.3914,-0.7142 & 222.3878,-0.7173 & $111\,(\pm 39)$ & 30 (0, 97) \\
HSCJ141105+002538 & 1.068 & 51.9 & 0.726/88 & 212.7725,0.4272 & 212.7777,0.4251 & 212.7733,0.4227 & $166\,(\pm30)$ & 136 (80, 192) \\
HSCJ161413+540413 & 0.872 & 47.2 & 0.736/94 & 243.5549,54.0702 & 243.5510,54.0731 & 243.5569,54.0714 & $105\,(\pm32)$ & 47 (0, 121) \\
HSCJ023018-062619 & 1.088 & 44.6 & 0.671/81 & 37.5753,-6.4387 & 37.5774,-6.4403 & 37.5778,-6.4432 & $80\,(\pm39)$ & 153 (86, 220) \\
HSCJ145727+002816 & 0.873 & 44.1 & 0.712/91 & 224.3640,0.4711 & 224.3604,0.4707 & 224.3655,0.4751 & $102\,(\pm73)$ & 119 (58, 181) \\
HSCJ085056-000931 & 0.882 & 44.1 & 0.710/91 & 132.7330,-0.1585 & 132.7358,-0.1562 & 132.7331,-0.1582 & $104\,(\pm28)$ & 11 (0, 48) \\
HSCJ232619+003017 & 1.172 & 41.7 & 0.635/76 & 351.5782,0.5048 & 351.5771,0.4975 & 351.5753,0.5002 & $222\,(\pm31)$ & 165 (105, 225) \\
HSCJ222625+013124 & 0.831 & 41.5 & 0.700/91 & 336.6045,1.5234 & 336.6017,1.5264 & 336.6003,1.5259 & $115\,(\pm35)$ & 133 (79, 188) \\
HSCJ145817-000720 & 1.002 & 43.0 & 0.677/83 & 224.5699,-0.1222 & 224.5721,-0.1199 & 224.5693,-0.1203 & $92\,(\pm29)$ & 60 (23, 97) \\ \hline
 \end{tabular}}\label{tbl:sample}
 \begin{tabnote}
    $^{\mathrm{a}}$ Richness. $^{\mathrm{b}}$Centroid offset
   (see section \ref{subsec:peakoffset} for
   definition); the standard error is given in parentheses. $^{\mathrm{c}}$Peak offset (see section~\ref{subsec:peakoffset} for definition). 
   The error range is indicated in parenthesis (see text). 
 \end{tabnote}
\end{table*}

\section{X-ray data analysis}\label{sec:analysis}
\subsection{Data reduction}\label{subsec:reduction}
The XMM-Newton observations were reduced following the standard procedures described in \citet{Ota20}. We processed the data using the XMM-Newton Science Analysis System (SAS) version~18, performing flare screening, point-source detection, and estimation of the quiescent particle background (see also \cite{Miyaoka18}). To further mitigate contamination from faint point sources, we cross-referenced the 4XMM-DR9 serendipitous source catalog \citep{Webb20} and excluded identified sources from subsequent analysis.

\subsection{Image analysis}\label{subsec:image}
\begin{figure*}[htb]
 \begin{center}
 \includegraphics[width=0.24\linewidth]{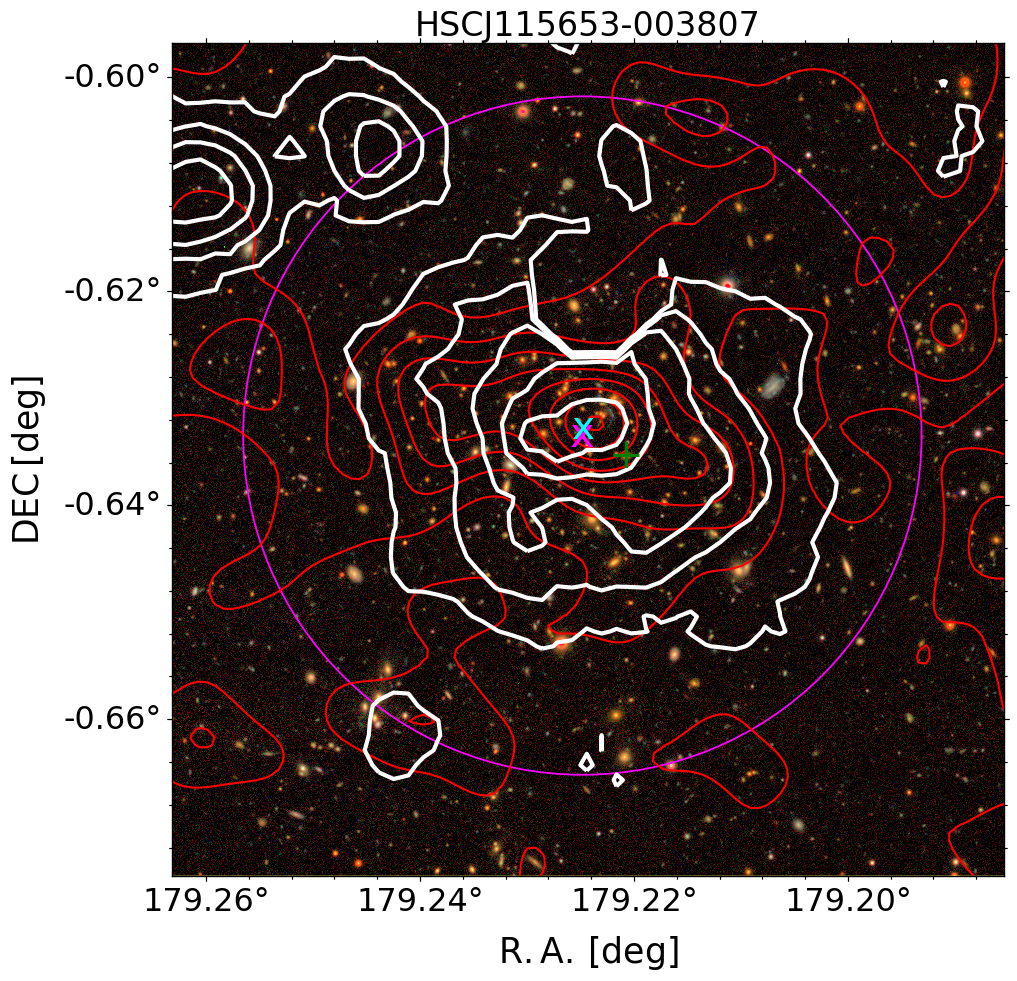} \includegraphics[width=0.24\linewidth]{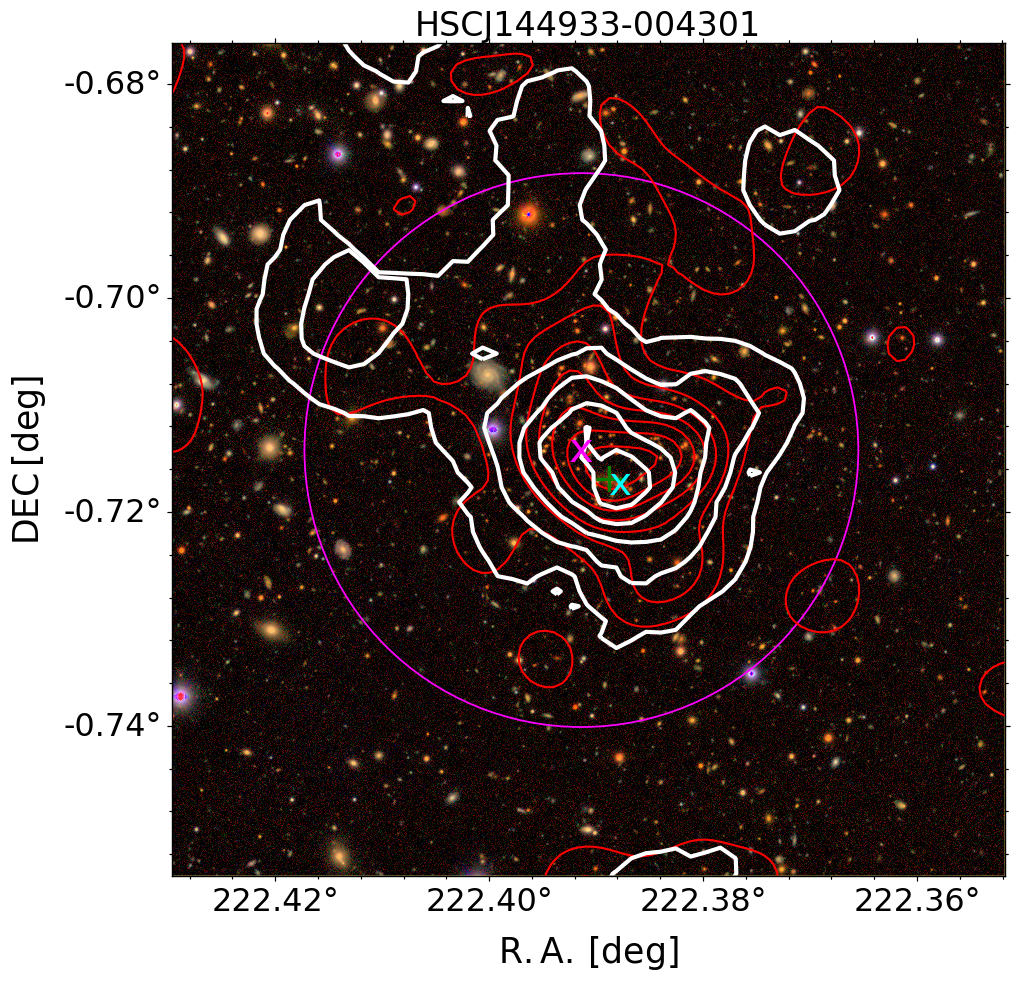} 
 \includegraphics[width=0.24\linewidth]{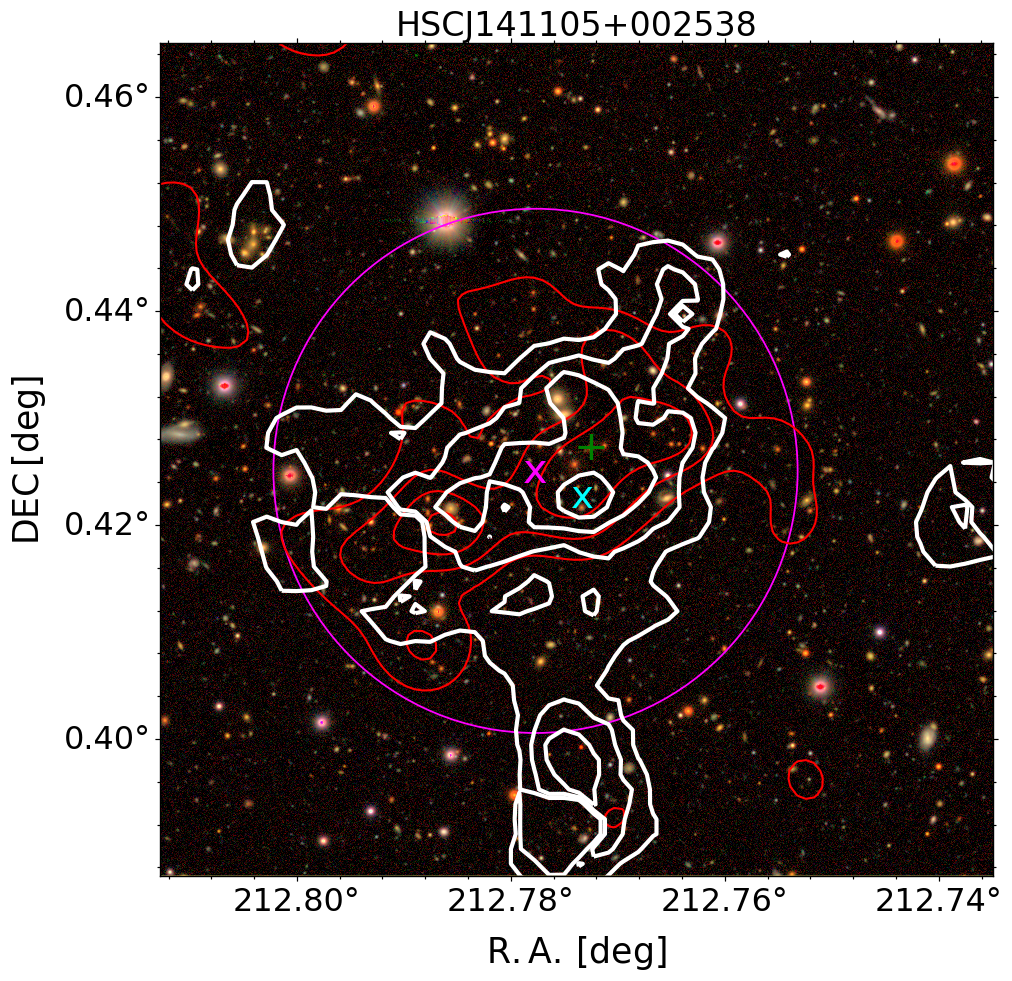} 
 \includegraphics[width=0.24\linewidth]{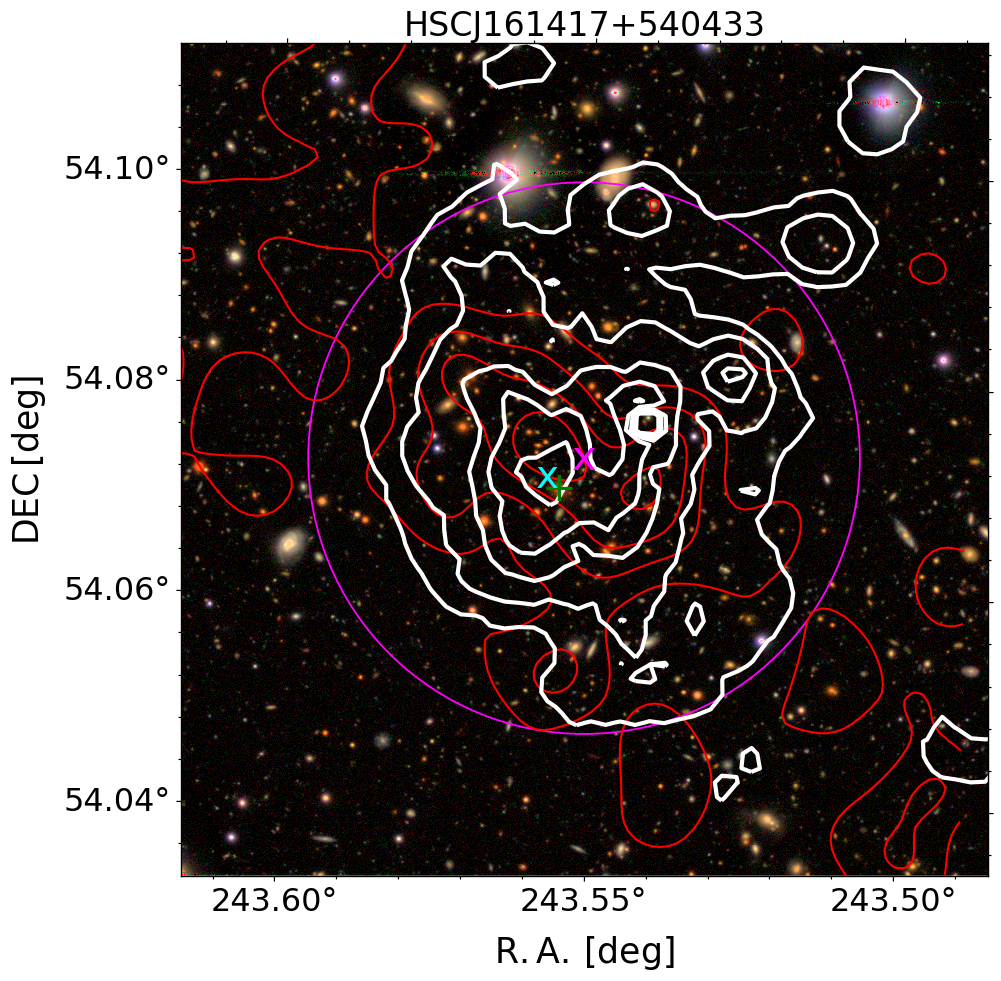} 
 \includegraphics[width=0.24\linewidth]{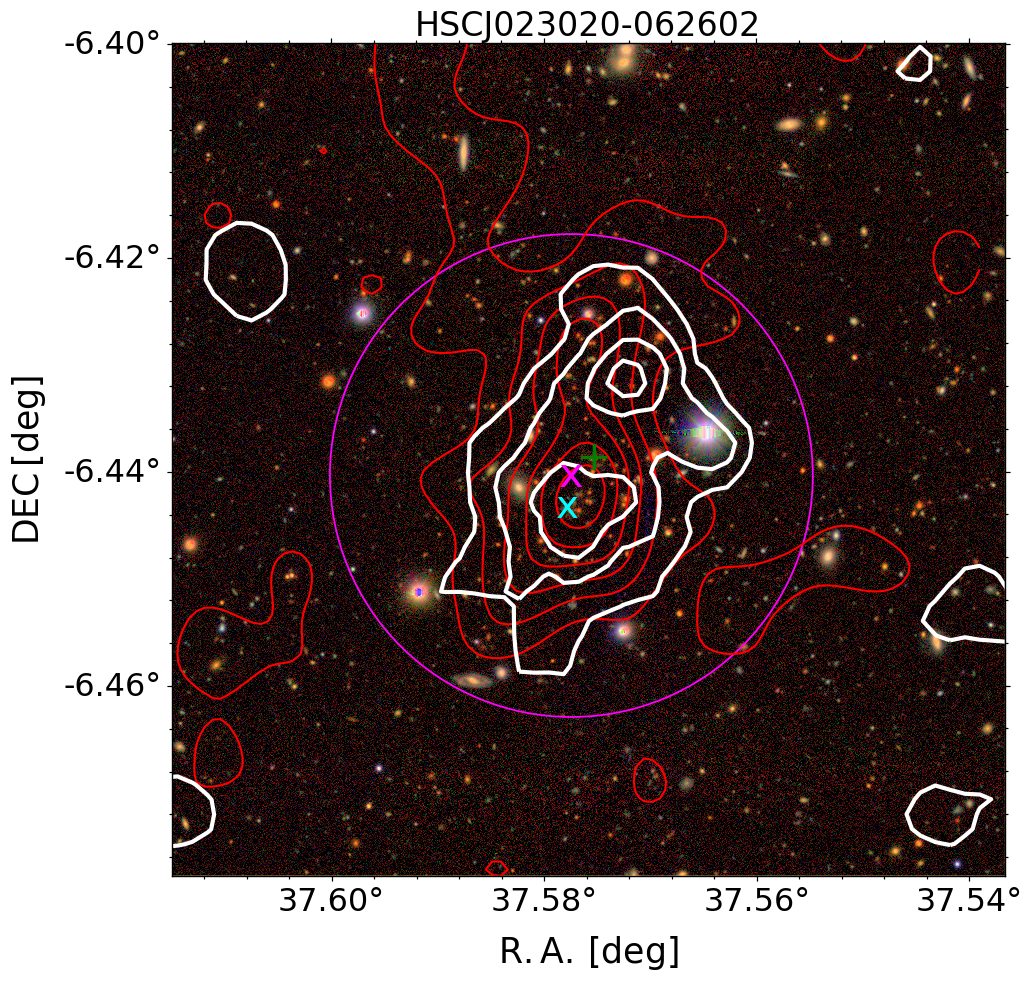} 
 \includegraphics[width=0.24\linewidth]{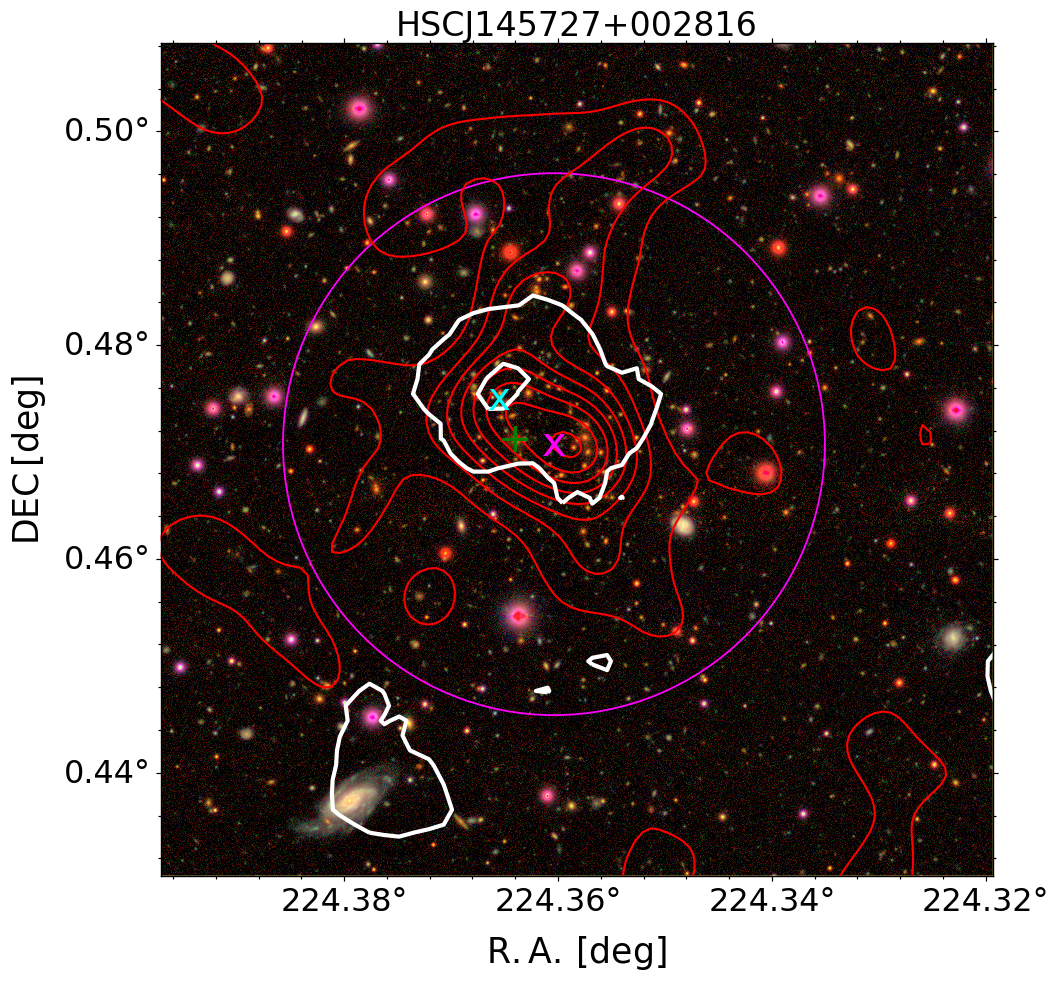} 
 \includegraphics[width=0.24\linewidth]{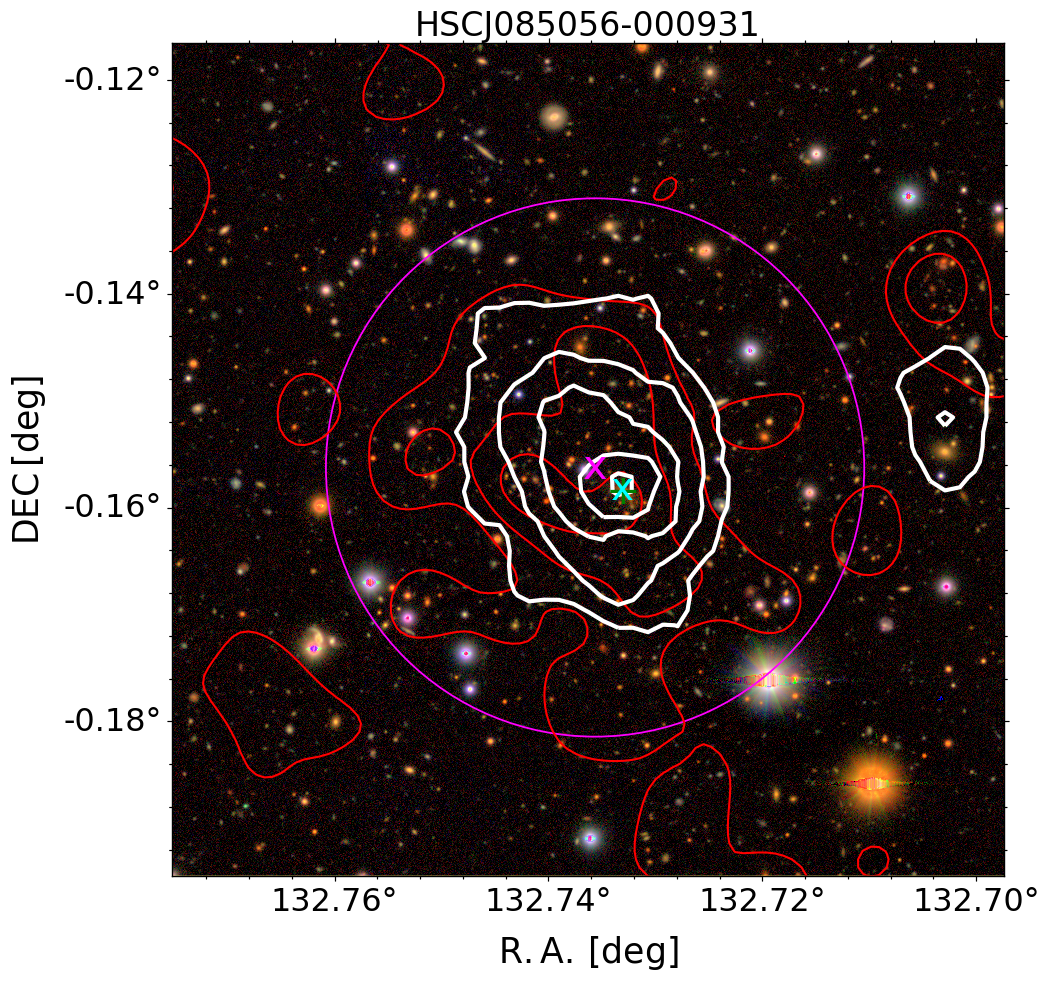} 
 \includegraphics[width=0.24\linewidth]{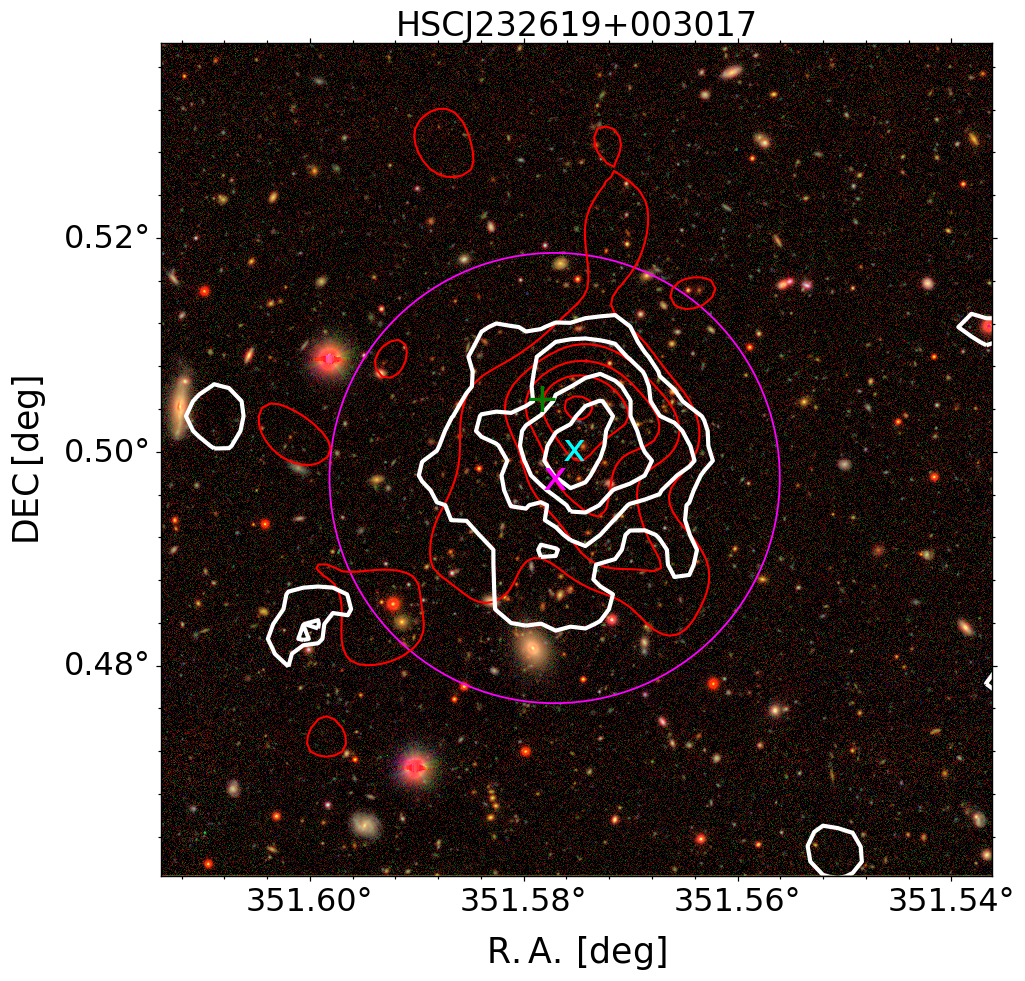} 
 \includegraphics[width=0.24\linewidth]{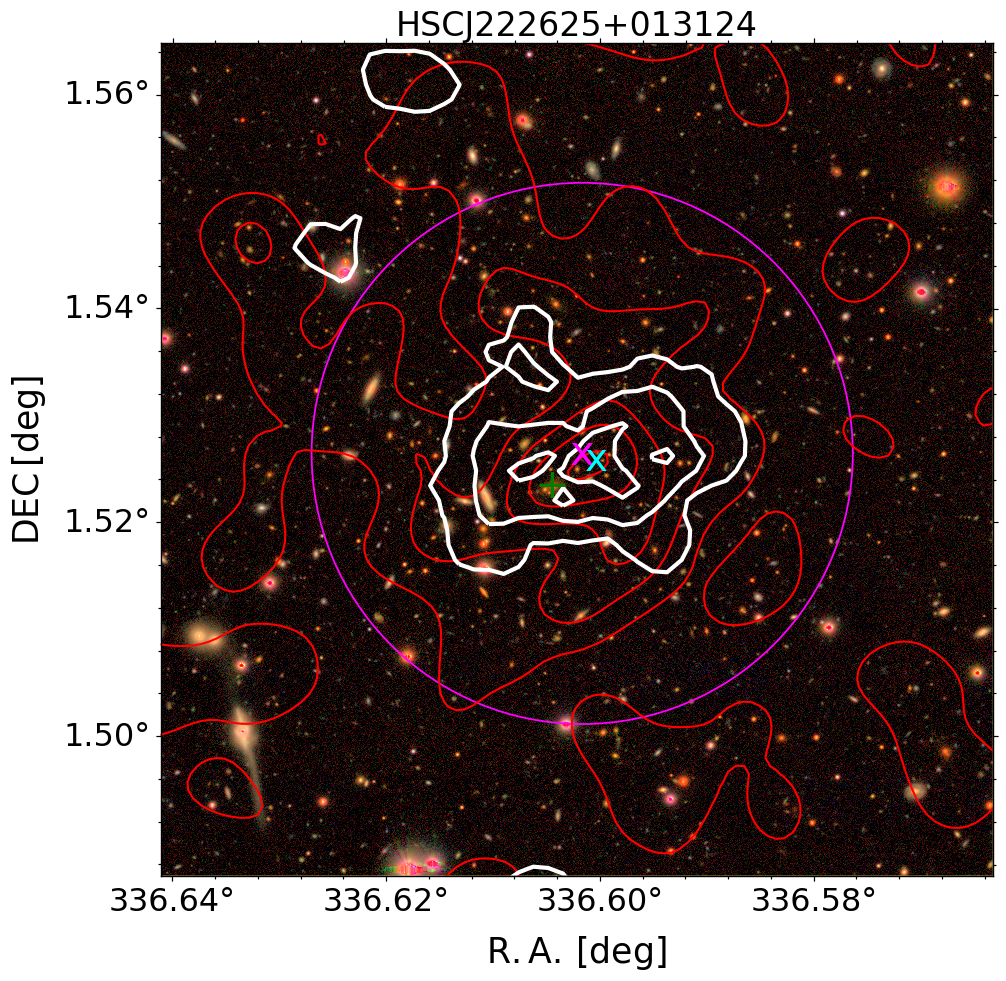} 
 \includegraphics[width=0.24\linewidth]{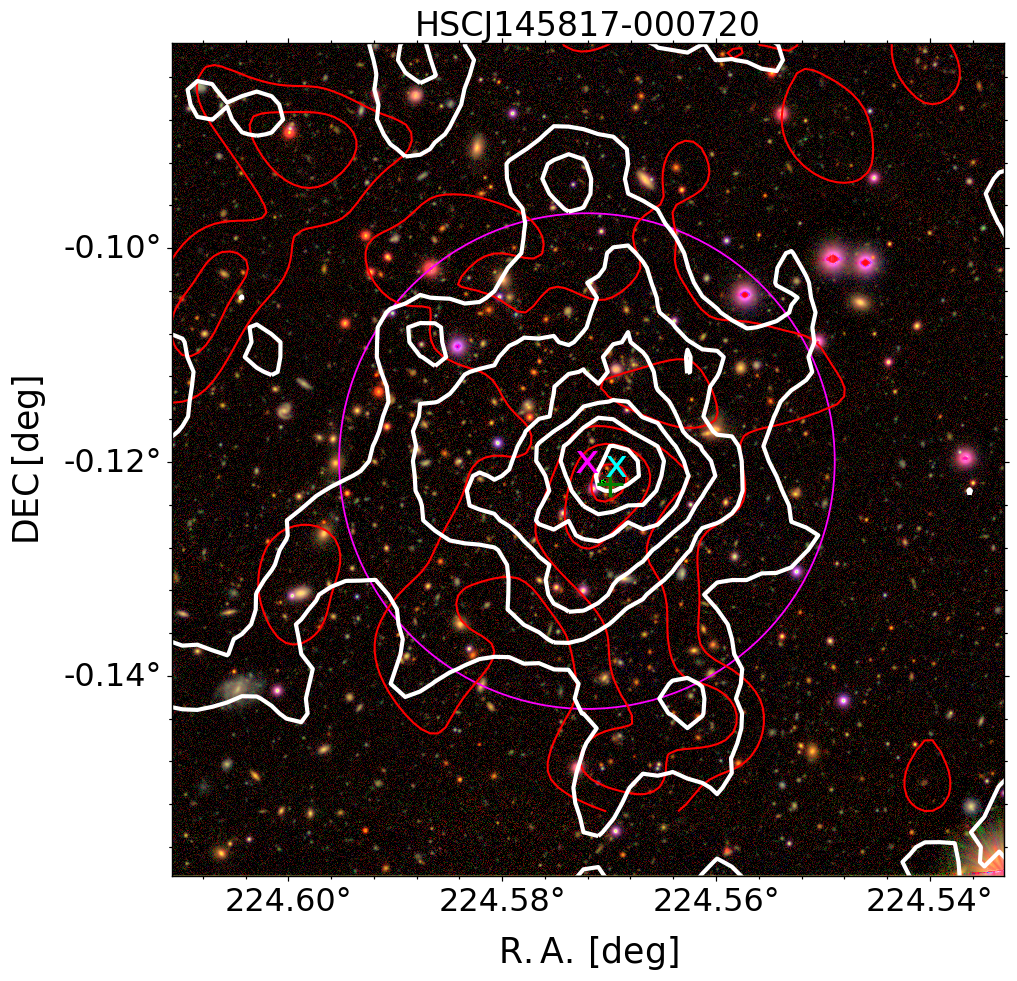} 
 \end{center}
 \caption{HSC I-band images of the high-redshift CAMIRA clusters (color) with superposed contours of X-ray intensities (white) and galaxy density maps (red). In each panel, the X-ray centroid, X-ray peak, and BCG positions are marked with a magenta ``x'', cyan ``+'', and green ``+'', respectively. 
 {Alt text: Multi panel figure showing optical images of galaxy clusters with overlaid maps. Each panel displays an I band image with contours indicating X ray surface brightness and contours indicating galaxy density. 
 }}\label{fig:image}
\end{figure*}

To characterize the spatial distribution of the ICM emission, we created background-subtracted and exposure-corrected EPIC composite images (MOS1, MOS2, and PN) in the 0.4--2.3~keV band, adopting a pixel scale of $5\arcsec$. Figure~\ref{fig:image} presents the HSC images of each cluster overlaid with X-ray brightness contours, clearly showing extended emission in all targets.

We first determined the X-ray centroid within a circular aperture of radius $R_{500}$. The value of $R_{500}$ was estimated using the $N-M_{500}$ relation of \citet{Okabe19} and the definition $M_{500} = (4\pi/3) R_{500}^3 \Delta_{c} \rho_{\rm crit}$. Here, $M_{500}$ is the total mass enclosed within $R_{500}$, which is defined as the radius within which the mean density is $\Delta_c = 500$ times the critical density of the Universe, $\rho_{\rm crit}$.    The resulting centroid position was subsequently adopted as the center for source spectrum extraction.

We then identified the X-ray peak within $R_{500}$ using a Gaussian-smoothed ($\sigma = 3$~pixels) EPIC image. The peak position is used to assess the cluster dynamical state via the offset between the BCG and X-ray peak. The systematic uncertainty of the peak location was evaluated through two tests: (i) varying the smoothing scale between 2, 3, and 4 pixels, and (ii) performing 1,000 image simulations with the {\tt SIXTE} simulator version 2.7.1 \citep{Dauser19}, assuming a symmetric $\beta$-model \citep{Cavaliere76} and constant background. The standard deviations of the recovered centroid and peak positions were adopted as estimates of the positional uncertainty.

\subsection{Spectral analysis}\label{subsec:spec}
To measure the ICM temperature and bolometric luminosity, we extracted spectra from a circular region of radius $R_{500}$ centered on the X-ray centroid, and background spectra from an annulus spanning $2R_{500}$ to $3R_{500}$. Corresponding response matrix files (RMFs) and ancillary response files (ARFs) were generated for each region.

The cluster emission was modeled with an absorbed APEC component \citep{Smith01, Foster12}. Hydrogen column densities were fixed to the HI4PI values \citep{HI4PI16}, and the metallicity was fixed at 0.3~solar, which lies within the range of average ICM abundances ($\sim0.2-0.4$~solar) reported for massive clusters up to $z\sim1.5$ \citep{Balestra07,McDonald16,Liu20}. At the redshift and photon statistics of our sample, the metallicity cannot be reliably constrained. We verified that varying the abundance between 0.2 and 0.4~solar changes the best-fit temperatures and bolometric luminosities by $\lesssim1\%$ and $\lesssim2\%$, respectively, which are significantly smaller than the statistical uncertainties. 

The background model consisted of five components: instrumental and solar charge-exchange lines (modeled with Gaussians), the Galactic X-ray background, the cosmic X-ray background, and residual soft proton contamination. For HSCJ145727+002816, the limited photon counts prevented a stable temperature determination; therefore, we fixed the temperature to the value predicted by the $N-T$ relation of \citet{Oguri18}.

\begin{table*}[hbt]
  \tbl{Results of X-ray and lensing data analysis}{%
  \begin{tabular}{llllllll}\hline\hline
  Cluster & OBSID$^{\mathrm{a}}$ & Exposure$^{\mathrm{b}}$ & $N_{\rm H}$ & $kT$ & $L_X$ & $\chi^2$/d.o.f.  & $M_{500}$ \\
          & & M1,M2,PN & ($10^{20}~{\rm cm^{-2}}$) & (keV) & ($10^{44}~{\rm erg\,s^{-1}}$) &  & ($10^{14}h_{70}^{-1}M_{\odot}$)\\ \hline
HSCJ115653-003807 & 0800400101 & 10.6,16.7,4.2 & 2.13 & $  4.15_{ -0.97}^{+0.89} $ & $  4.80_{ -0.95}^{+0.62} $ & 178.4/138 & $  3.04_{ -2.46}^{+2.78} $ \\ 
HSCJ144933-004301 & 0800400301 & 10.2,12.8,6.6 & 3.93 & $  4.14_{ -1.41}^{+1.90} $ & $  5.00_{ -1.66}^{+0.75} $ & 104.5/90 & $  6.79_{ -4.34}^{+5.55} $ \\ 
HSCJ141105+002538 & 0800400201 & 28.4,28.8,24.6 & 2.74 & $  4.36_{ -0.76}^{+0.94} $ & $  4.71_{ -0.49}^{+0.50} $ & 276.9/263 & $  8.09_{ -5.00}^{+6.33} $ \\ 
HSCJ161413+540413 & 0821290201 & 19.6,21.2,12.0 & 1.05 & $  7.23_{ -1.38}^{+1.12} $ & $  7.46_{ -1.40}^{+0.63} $ & 227.5/203 & -- \\ 
HSCJ023018-062619 & 0800400401 & 22.6,23.8,13.9 & 2.53 & $  4.61_{ -1.19}^{+1.42} $ & $  3.84_{ -0.77}^{+0.94} $ & 159.6/121 & -- \\ 
HSCJ145727+002816 & 0821290501 & 11.2,12.0,7.8 & 3.97 & $  3.66_{ -0.16}^{+0.17} $ & $  0.50_{ -0.35}^{+0.22} $ & 151.6/88 & $  4.27_{ -2.66}^{+3.36} $ \\ 
HSCJ085056-000931 & 0800400601 & 23.2,24.2,12.3 & 2.76 & $  3.12_{ -0.51}^{+0.53} $ & $  3.42_{ -0.34}^{+0.31} $ & 186.2/145 & $  3.78_{ -3.43}^{+3.43} $ \\ 
HSCJ232619+003017 & 0821290601 & 36.6,36.8,29.3 & 3.93 & $  4.52_{ -0.79}^{+0.80} $ & $  4.32_{ -0.56}^{+0.60} $ & 249.1/224 & $  3.81_{ -7.03}^{+8.33} $ \\ 
HSCJ222625+013124 & 0800400901 & 28.2,29.0,19.0 & 4.76 & $  5.04_{ -1.50}^{+2.47} $ & $  2.26_{ -0.48}^{+0.41} $ & 274.5/236 & $  4.81_{ -2.65}^{+3.32} $ \\ 
HSCJ145817-000720 & 0821290401 & 11.8,12.0,18.4 & 4.11 & $  5.08_{ -0.85}^{+0.57} $ & $  8.38_{ -1.05}^{+0.50} $ & 193.5/166 & $  4.17_{ -5.60}^{+6.88} $ \\ \hline
 \end{tabular}}\label{tbl:spec}
  \begin{tabnote}
  $^{\mathrm{a}}$The XMM-Newton observation id. $^{\mathrm{b}}$ The XMM-Newton EPIC-MOS1(M1), MOS2(M2), and PN exposure time after data filtering (ksec).
   \end{tabnote}
\end{table*}

We performed simultaneous fits to the source and background spectra using XSPEC version~12.11.1 \citep{Arnaud96}, adopting energy bands of 0.3--10~keV for MOS and 0.4--10~keV for PN. Best-fit APEC parameters are listed in table~\ref{tbl:spec}. Bolometric luminosities were derived from the best-fit model in the 0.01--30~keV band and corrected for flux lost due to point-source masking by interpolating the cluster surface brightness with a $\beta$-model.

\section{Lensing data analysis}\label{subsec:lens}
We derive weak-lensing (WL) masses for clusters located within the footprint of the HSC-Y3 shape catalog \citep{HSC3YShape}, which is based on galaxy shapes measured with the re-Gaussianization PSF-correction method \citep{Hirata03}. The catalog includes galaxies satisfying full-color and full-depth criteria, enabling accurate shape measurements and reliable photometric redshift estimates. Two clusters, HSCJ161413+540413 and HSCJ023018-062619, lie outside the HSC-Y3 coverage and are therefore excluded from the WL analysis.

Because our clusters are at high redshift, we apply the color--color selection method of \citet{Medezinski17} to identify background galaxies. We adopt the X-ray centroid of each cluster as the center for the WL mass measurement to ensure consistency with the X-ray analysis.

The WL mass measurement follows the method of \citet{Okabe25}, using an NFW halo model \citep{NFW96,NFW97} with the overdensity fixed at $\Delta_c=500$. Due to the limited number of usable background galaxies, the measured WL masses $M^{\rm WL}$ are subject to a positive bias: the lensing shear is estimated from the composite distortion of intrinsically elliptical galaxies, leading to statistical skewness in the inferred mass \citep{Okabe25}. To model this effect, we perform mock simulations assuming spherical NFW halos combined with realistic background number densities.

The WL mass bias is incorporated into the scaling-relation regression through the relation
\begin{equation}
    \ln x^{\rm WL} = a_{\rm WL} +  (b_{\rm WL} + d_{\rm WL}\ln E(z))\ln x  + c_{\rm WL} \ln E(z),
\end{equation}
where $x^{\rm WL}=M^{\rm WL}$ and $x=M$ or $x^{\rm WL}=M^{\rm WL}E(z)$ and $x=ME(z)$ with $E(z) = \left[ \Omega_{\mathrm{M}}(1+z)^3 + \Omega_{\Lambda} \right]^{1/2}$. Here, $M$ denotes the true underlying halo mass.  Intrinsic scatter is modeled as
\begin{equation}
\sigma_x=\sigma_{\rm WL}+\sigma_{{\rm WL},c}\ln E(z). 
\end{equation}
All the WL mass bias parameters are treated as priors with the error covariance matrix in regression analysis (section \ref{subsec:scaling}). The best-fit parameters and their diagonal errors are summarized in table~\ref{tab:Mbias}.

\begin{table}[h]
    \tbl{WL mass bias parameters}{
    \begin{tabular}{ccc}\hline\hline
   $x$      &  $M$ & $ME(z)$\\ \hline
    $a_{{\rm WL}}$ & $-0.020_{-0.042}^{+0.043}$ & $-0.033_{-0.046}^{+0.045}$\\
 $b_{{\rm WL}}$ & $1.004_{-0.020}^{+0.019}$ & $1.005_{-0.020}^{+0.020}$ \\
 $c_{{\rm WL}}$ & $1.009_{-0.263}^{+0.254}$  & $1.196_{-0.294}^{+0.287}$\\
 $d_{{\rm WL}}$ & $-0.399_{-0.124}^{+0.125}$ & $-0.408_{-0.125}^{+0.127}$\\
 $\sigma_{{\rm WL}}$ & $0.033_{-0.003}^{+0.003}$  & $0.032_{-0.003}^{+0.003}$ \\
 ${\sigma_{{\rm WL},c}}$ & $-2.059_{-0.023}^{+0.024}$  & $-2.073_{-0.023}^{+0.024}$  \\ \hline
    \end{tabular}} \label{tab:Mbias}
\end{table}

\section{Results} 
\subsection{X-ray peak offset and morphological classification}\label{subsec:peakoffset}
We assessed the dynamical states of the clusters by measuring the projected offsets between the BCG and both the X-ray centroid ($D_{\mathrm{XC}}$) and the X-ray peak ($D_{\mathrm{XP}}$) within $R_{500}$. Individual measurements are listed in table~\ref{tbl:sample}. Typical uncertainties on $D_{\mathrm{XC}}$ and $D_{\mathrm{XP}}$ were derived from variations in the smoothing kernel and from image simulations conducted with {\tt SIXTE} (see section~\ref{subsec:image}).

\begin{figure*}[hbt]
\begin{center}
\includegraphics[width=0.4\linewidth]{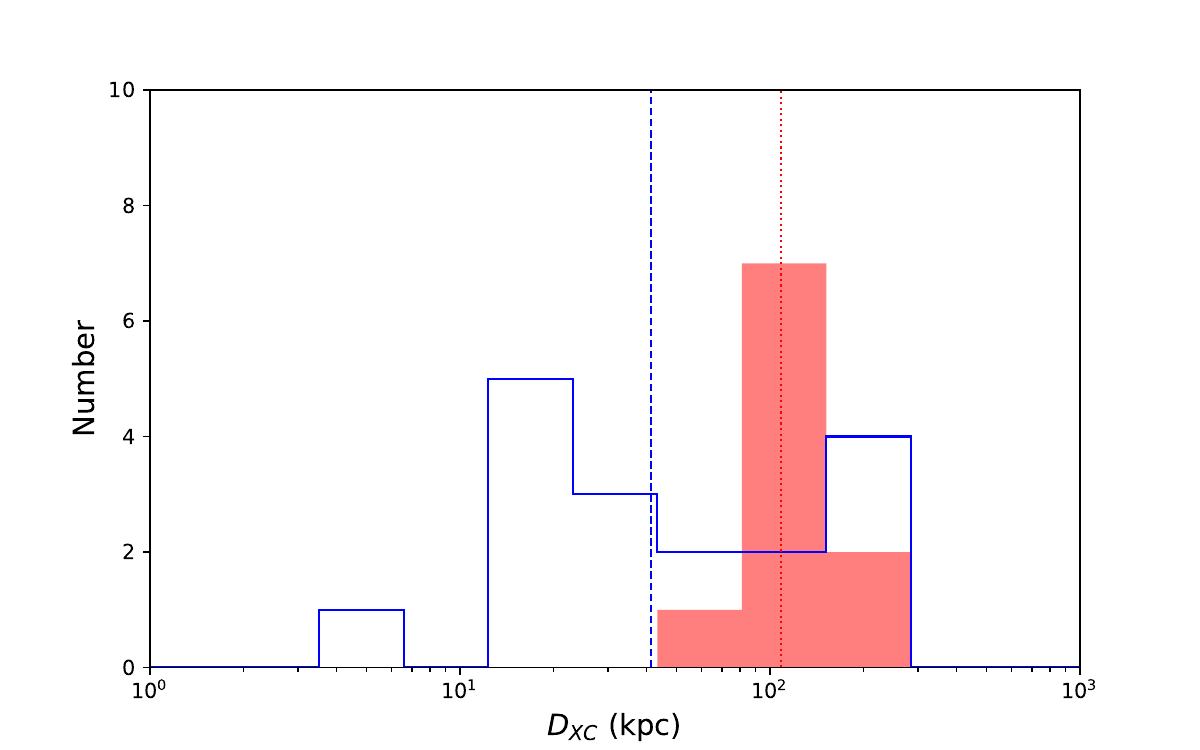}
\includegraphics[width=0.4\linewidth]{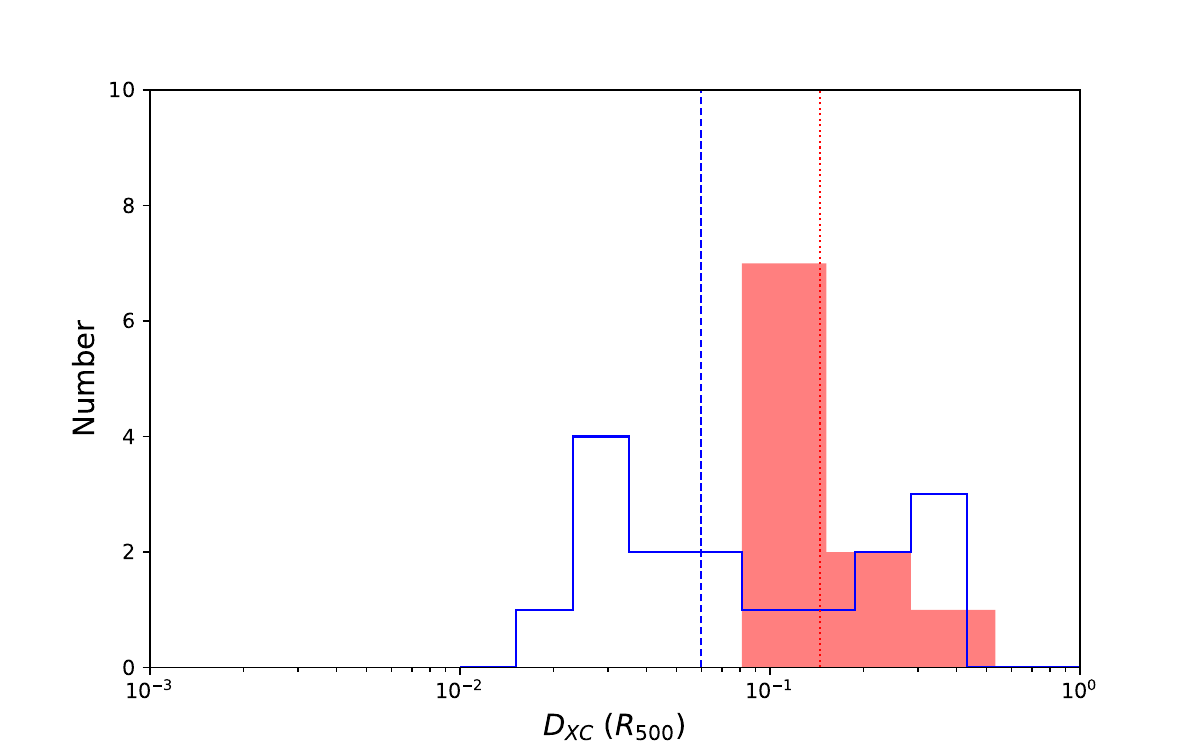}
\includegraphics[width=0.4\linewidth]{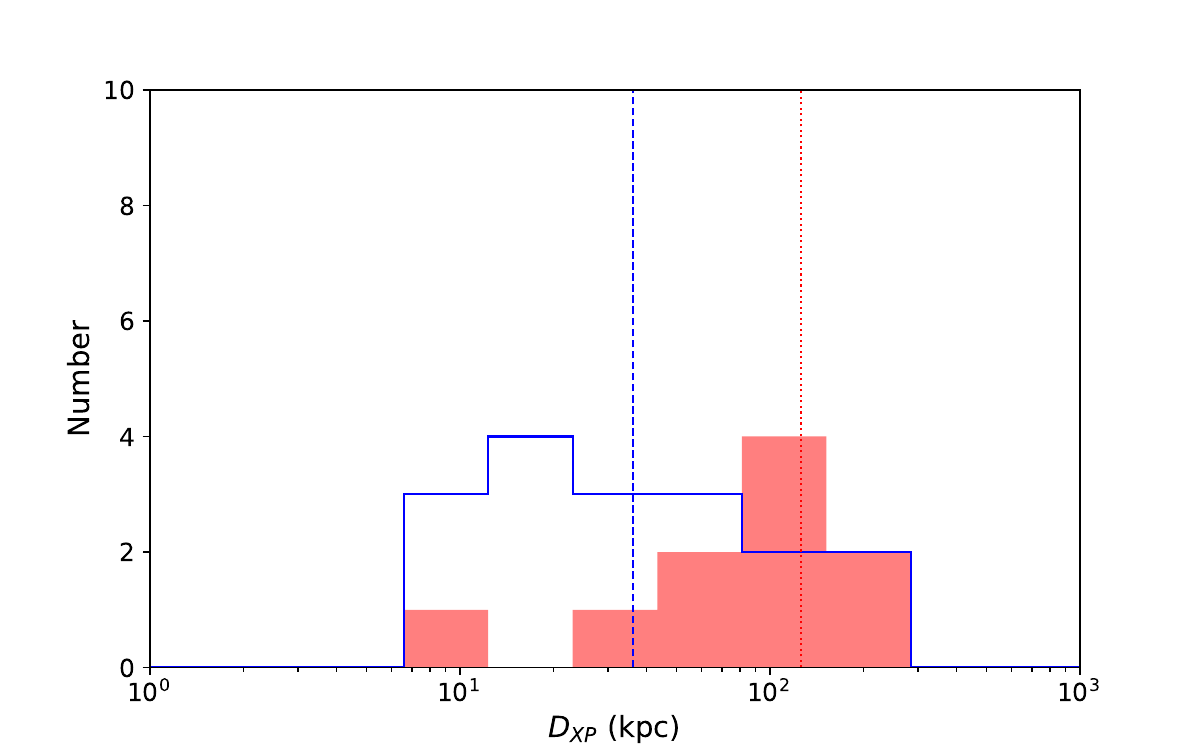}
\includegraphics[width=0.4\linewidth]{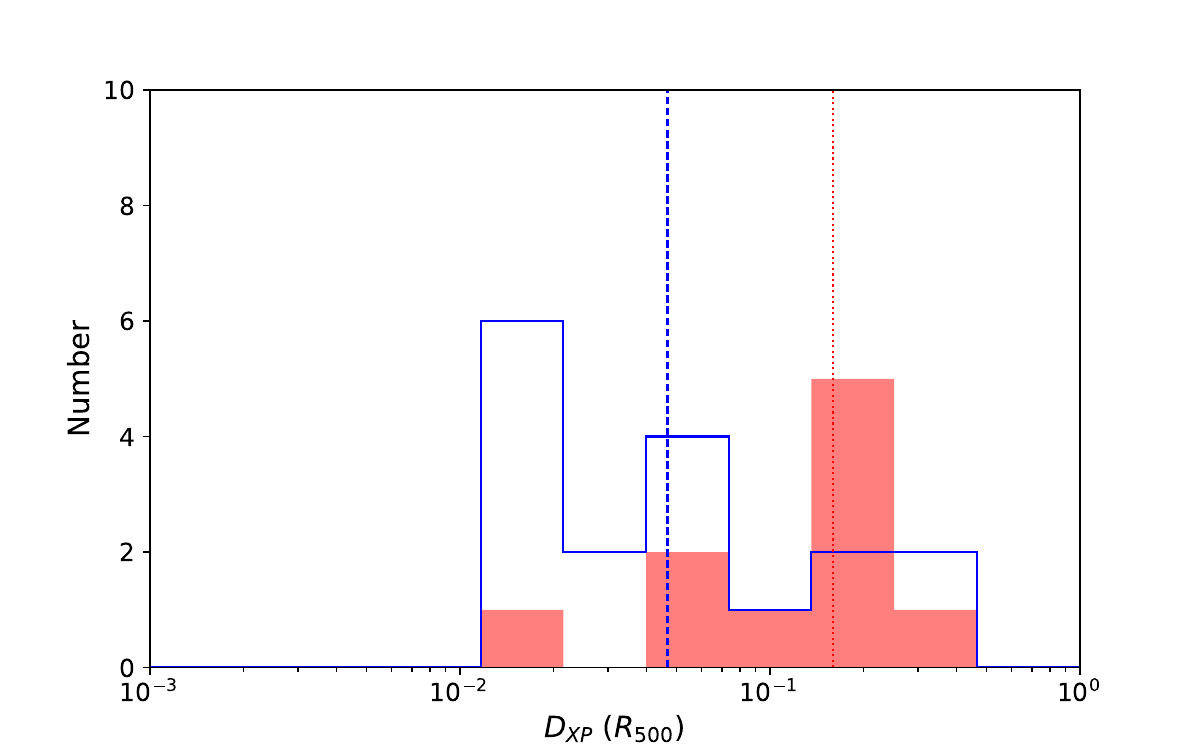}
\end{center}
\caption{Histograms of BCG-X-ray centroid offset and BCG-X-ray peak offset in units of kpc and $R_{500}$. The filled and open histograms show the distributions of 10 high-redshift clusters (this work) and 17 low-redshift clusters \citep{Ota20}, respectively. The vertical dotted line and dashed line indicate the median of two samples. 
{Alt text: Multiple histograms comparing distributions of brightest cluster galaxy offsets from the X ray centroid and from the X ray peak. Each histogram is presented both in kiloparsec units and normalized by the cluster radius R five hundred. }} \label{fig:bcg_offset}
\end{figure*}

Figure~\ref{fig:bcg_offset} presents the distributions of $D_{\mathrm{XC}}$ and $D_{\mathrm{XP}}$ in physical units (kpc) and normalized by $R_{500}$, comparing our high-redshift sample with the low-redshift CAMIRA clusters of \citet{Ota20}. The high-$z$ clusters exhibit significantly larger offsets, with median values of 108~kpc ($D_{\mathrm{XC}}$) and 126~kpc ($D_{\mathrm{XP}}$), compared to 41~kpc and 36~kpc for the low-$z$ sample.

Following the morphological classification criterion of \citet{Sanderson09}, we define clusters with $D_{\mathrm{XP}} < 0.02 R_{500}$ as dynamically ``relaxed.'' Only one of the ten high-redshift clusters satisfies this condition, yielding a relaxed fraction of 10\%. The statistical and systematic uncertainties are estimated to be 9\% and 10\%, respectively (table~\ref{tbl:bcg_offset}). Image simulations provide a comparable total uncertainty of $\sim 20\%$, implying an upper limit of $\sim 30\%$ on the relaxed fraction. These results indicate that high-redshift CAMIRA clusters are, on average, dynamically disturbed systems.

\begin{table}[hbt]
  \tbl{Results of BCG-X-ray offset measurements}{%
  \begin{tabular}{llllll}\hline\hline
    & $D_{\rm XC}$ & $D_{\rm XC}$ & $D_{\rm XP}$ & $D_{\rm XP}$ & Fraction$^{\mathrm{a}}$ \\ 
       & (kpc) & ($R_{500}$) & (kpc) & ($R_{500}$) &   \\ \hline
High-z$^{\mathrm{b}}$ & 108 & 0.15 & 126 & 0.16 & $10\pm 9 \,(\pm 10) $ \%  \\
Low-z$^{\mathrm{c}}$ & 41  & 0.06 & 36  & 0.05 & $29\pm 11 \,(\pm 13)$ \% \\ 
 \hline
  \end{tabular}}\label{tbl:bcg_offset}
  \begin{tabnote}
 $^{\mathrm{a}}$ Fraction of relaxed clusters. The first and second errors are the statistical and systematic uncertainties, respectively. $^{\mathrm{b}}$ This work. $^{\mathrm{c}}$ \citet{Ota20}. 
 \end{tabnote}
\end{table}

\subsection{Scaling relations}\label{subsec:scaling}
To investigate scaling relations and their possible redshift evolution, we combined our ten high-$z$ clusters with 17 low-$z$ CAMIRA clusters from \citet{Ota20}. We employed the hierarchical Bayesian regression code HiBRECS \citep{Akino22}, which fits two observables ($y_1$, $y_2$) as power-law functions of a third variable $x$ while accounting for intrinsic scatter and redshift evolution through the relation
\begin{equation}
\log{y_i}  =  a_i + b_i\log{x} + c_i\log{E(z)}, \label{eq:scaling}
\end{equation}
We performed two sets of joint fits, simultaneously fitting the $N$--$T$ and $L$--$T$ relations in one case, and the $N$--$M$ and $L$--$M$ relations in the other. The intrinsic scatter in WL mass was fixed to $\ln \sigma_M = 1.54$ following \citet{Umetsu20}. We evaluated three evolutionary models: (i) no redshift evolution ($c = 0$), (ii) free evolution ($c$ free), and (iii) self-similar evolution based on \citet{Kaiser86}, where $E(z)^{-1}L \propto T^2$ and $E(z)^{-1}L \propto [E(z)M]^{4/3}$.

\begin{figure*}[htb]
\begin{center}
\includegraphics[width=0.45\linewidth]{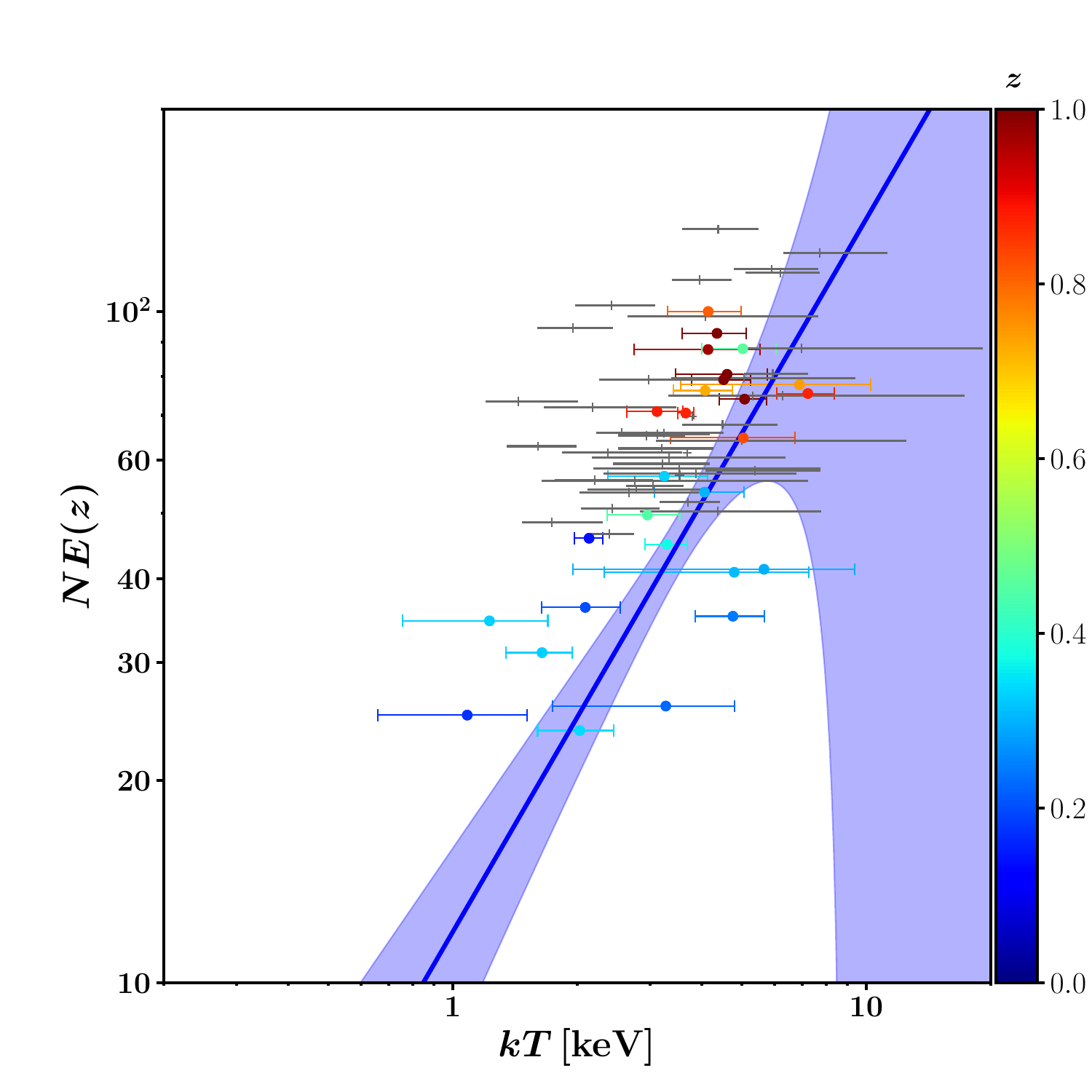}
\includegraphics[width=0.45\linewidth]{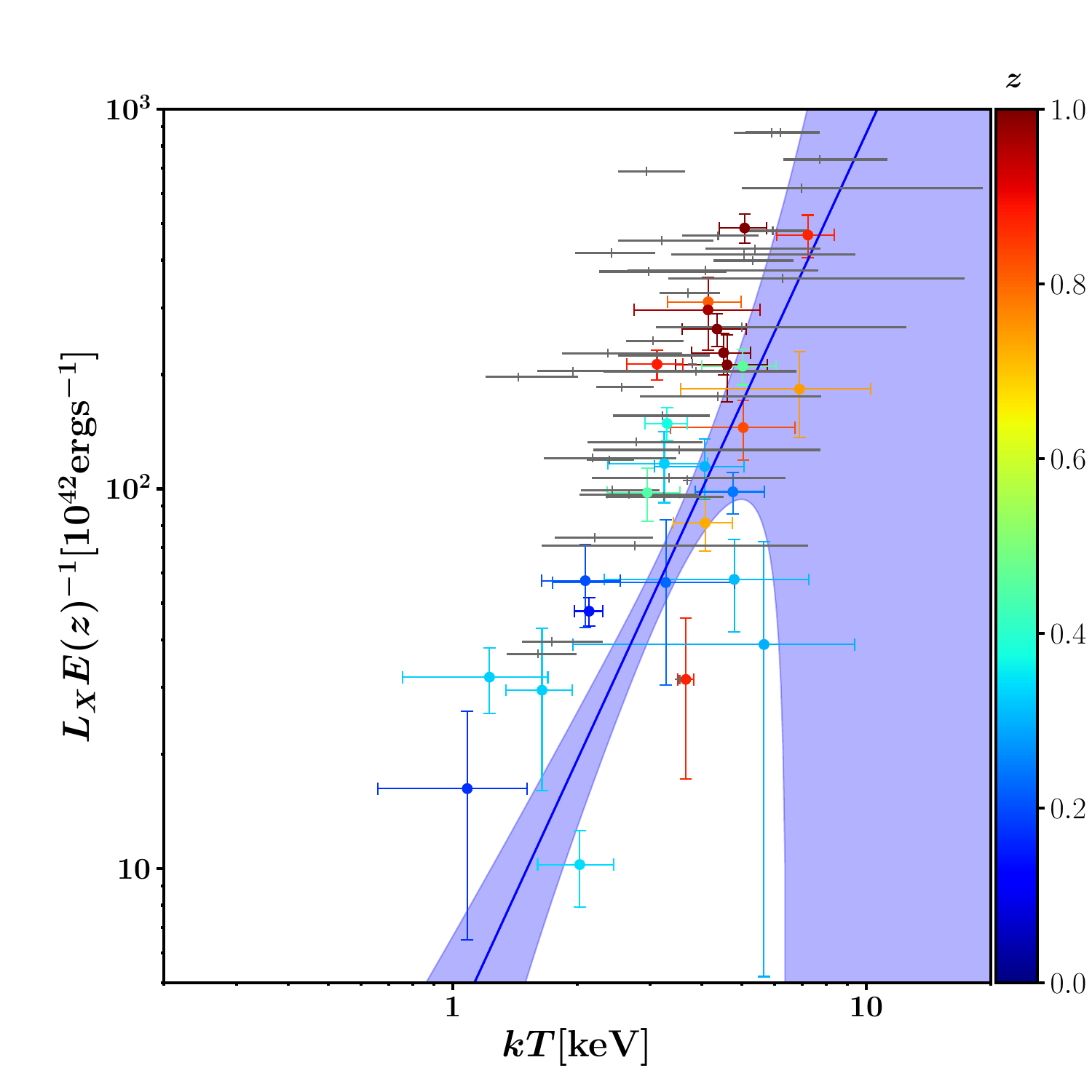}
\includegraphics[width=0.45\linewidth]{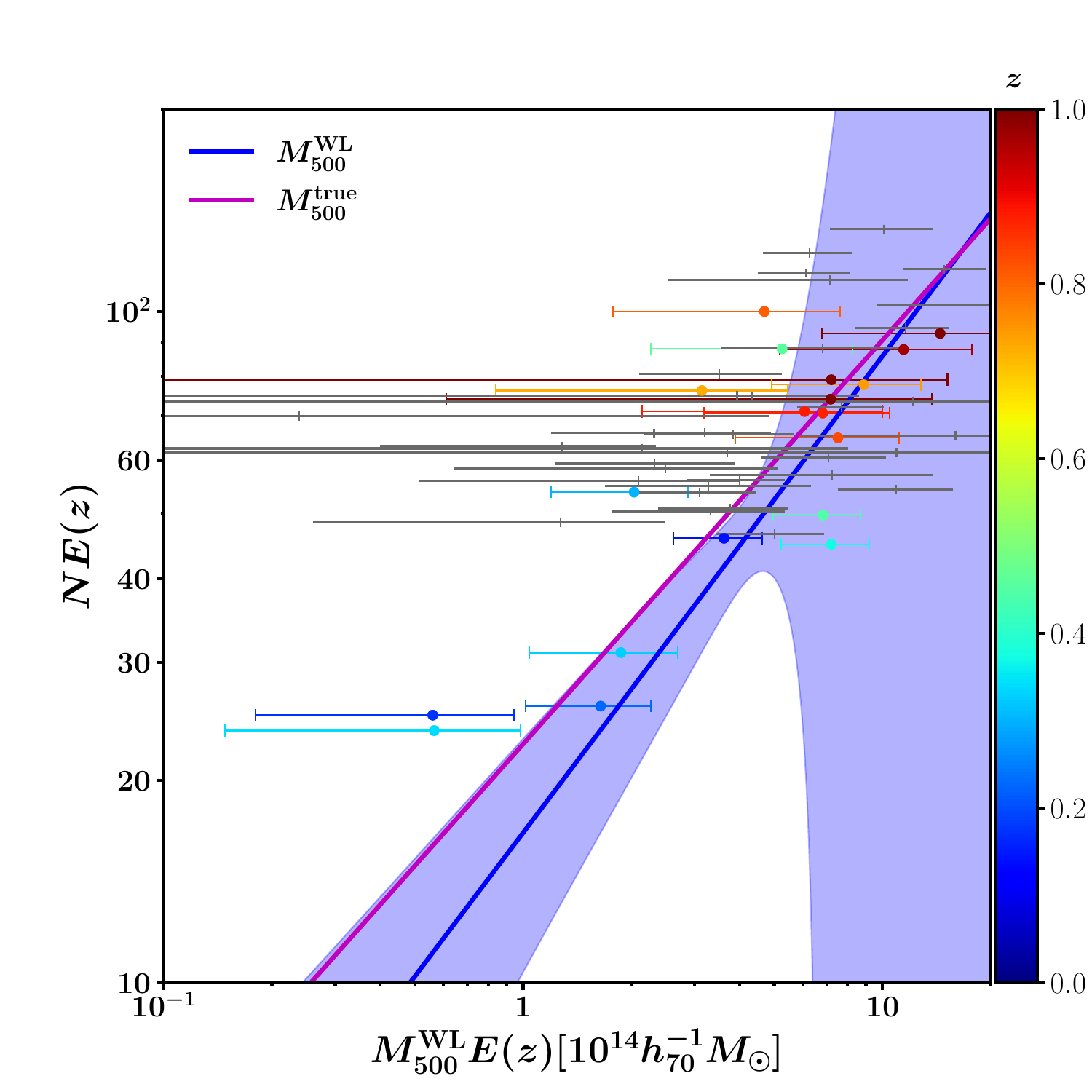}
\includegraphics[width=0.45\linewidth]{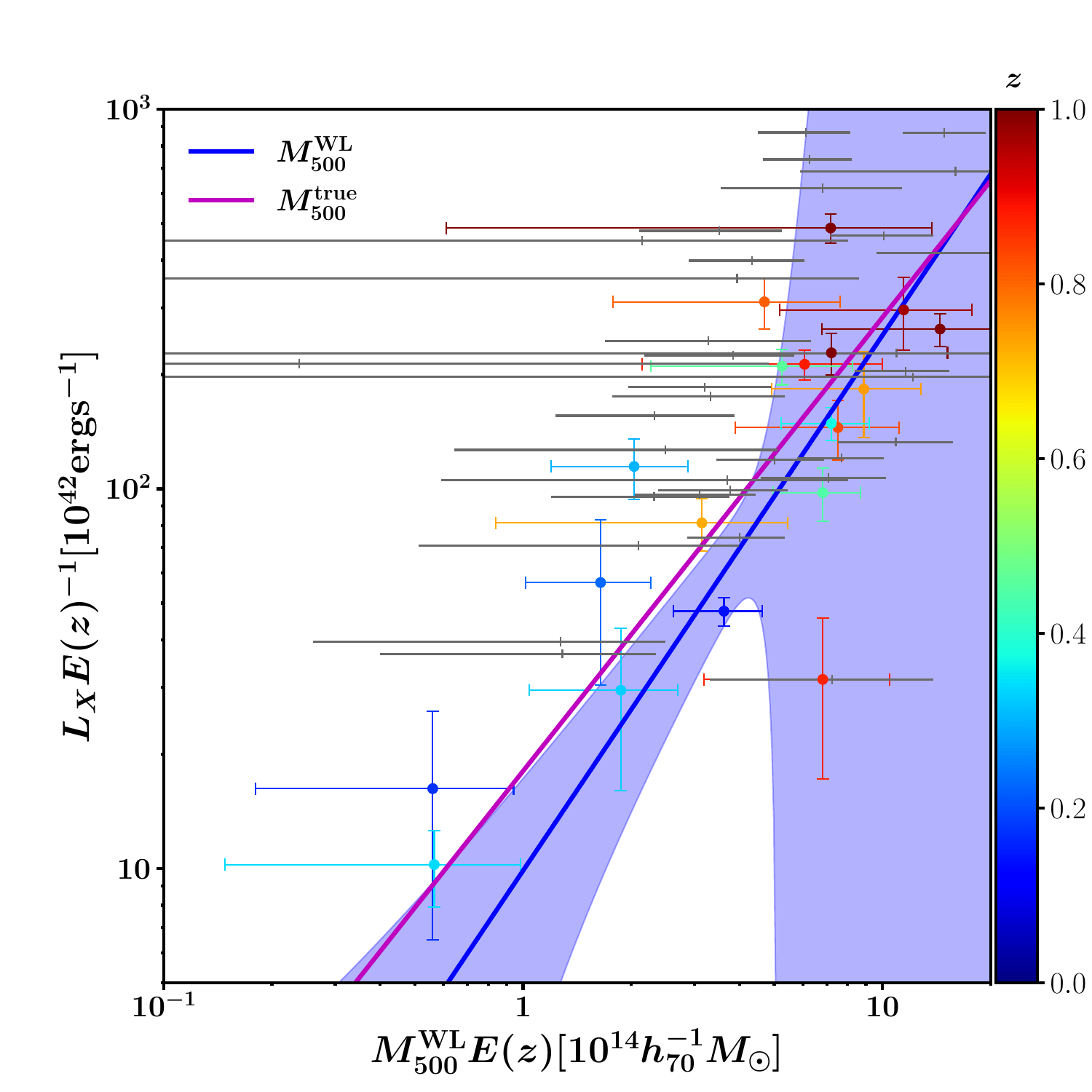}
\end{center}
\caption{Scaling relations of the CAMIRA clusters. The upper two panels show the richness--temperature ($N$--$T$) and luminosity--temperature ($L$--$T$) relations, while the lower two panels show the richness--mass ($N$--$M$) and luminosity--mass ($L$--$M$) relations. Each data point is color-coded according to the redshift of the cluster, as indicated by the color bar. For comparison, high-richness CAMIRA clusters observed in the eFEDS field \citep{Ota23} are shown in gray. The solid lines represent the best-fit scaling relations, calculated for the reference redshift $z=0.79$.
{Alt text: Multi panel scatter plots showing scaling relations for the CAMIRA clusters. Each panel displays data points plotted against fitted trend lines. The points include two samples: CAMIRA clusters in the present study and comparison clusters from the eFEDS field shown as separate markers.}}\label{fig:scaling}
\end{figure*}

The best-fit parameters and Akaike Information Criterion (AIC; \cite{Akaike74}) values for each model are listed in table~\ref{tbl:scaling}. The $\Delta$AIC values represent differences relative to the no-evolution model. Figure~\ref{fig:scaling} shows the scaling relations for the self-similar case.

Comparing models (i) and (ii), allowing $c$ to vary yields a slightly lower AIC for the $L-M$ relation, while the $L-T$ relation remains nearly unchanged. However, the large uncertainties on $c$ indicate that the improvement is not statistically meaningful. Under the self-similar assumption (model iii), the AIC values for the $L-M$ relation are comparable to those of the other two models, whereas the $L-T$ relation exhibits a marginally higher AIC. Despite this, the resulting slopes and normalizations remain consistent within uncertainties across all models.

Overall, these results suggest that the inclusion of a redshift-evolution term provides at most a marginal improvement to the fits. Given the large uncertainties and small AIC differences, the present data do not strongly support significant redshift evolution beyond the self-similar expectation.

\begin{table*}[hbt]
  \tbl{Best-fit scaling relations of the CAMIRA clusters}{
  \begin{tabular}{lllllllllllll}\hline\hline
   &  &       &       & \multicolumn{4}{c}{$y_1-x$ relation}      & \multicolumn{4}{c}{$y_2-x$ relation}  & \\  \cline{5-8}\cline{9-12}\
   & $x$ & $y_1$ & $y_2$ & $a_1$ & $b_1$ & $c_1$ & $\sigma_1$ & $a_2$ & $b_2$ & $c_2$ & $\sigma_2$  & $\Delta$AIC \\ \hline
(i) & $T$ & $N$ & $L$ & $ 1.34 \pm  0.11$ & $ 0.42 \pm  0.22$ & 0.00 (fix) & 0.24 & $ 0.97 \pm  0.32$ & $ 2.29 \pm  0.79$ & 0.00 (fix)  & 0.24 & 0 \\ 
(ii) & $T$ & $N$ & $L$ & $ 1.23 \pm  0.16$ & $ 0.79 \pm  0.40$ & $-0.33 \pm  1.24$ & 0.25 & $ 0.94 \pm  0.23$ & $ 2.09 \pm  0.51$ &  $ 1.89 \pm  1.97$ & 0.08 & 0.6 \\ 
(iii) & $T$ & $NE(z)$ & $LE(z)^{-1}$ &  $ 1.27 \pm  0.10$ & $ 0.90 \pm  0.22$ & 0.00 (fix) &  0.19 &  $ 1.01 \pm  0.20$ & $ 2.01 \pm  0.46$ & 0.00 (fix) & 0.34 & 6.6 \\ \hline  
(i) & $M$ & $N$ & $L$ &  $ 1.34 \pm  0.11$ & $ 0.41 \pm  0.19$ & 0.00 (fix) & 0.26 & $ 1.26 \pm  0.26$ &  $ 1.61 \pm  0.48$ & 0.00 (fix) & 0.09 & 0 \\ 
(ii) & $M$ & $N$ & $L$ &  $ 1.22 \pm  0.29$ & $ 0.74 \pm  0.51$ & $-0.34 \pm  1.23$ &  0.26 & $ 1.06 \pm  0.44$ & $ 1.35 \pm  0.85$ & $ 3.44 \pm  1.50$ & 0.43 & $-1.7$ \\ 
(iii) & $ME(z)$ & $NE(z)$ & $LE(z)^{-1}$ & $ 1.36 \pm  0.13$ &  $ 0.60 \pm  0.21$ & 0.00 (fix) &  0.18 &  $ 1.20 \pm  0.27$ & $ 1.19 \pm  0.41$ & 0.33 (fix) & 0.41 & $-1.2$ \\ \hline  
\end{tabular}}\label{tbl:scaling}
\begin{tabnote}
The best-fit parameters $a_i$, $b_i$, and $c_i$ in equation~\ref{eq:scaling}, as well as the intrinsic scatter $\sigma_i$, obtained from the hierarchical Bayesian fits are listed.
\end{tabnote}
\end{table*}

\section{Discussion}
In this section, we interpret the main observational results in the context of high-redshift cluster evolution. We first examine the dynamical states inferred from the BCG-X-ray offsets, then evaluate the redshift dependence of the ICM scaling relations, and finally discuss the elevated AGN activity and its potential impact on the thermodynamic state of the ICM.

\subsection{Cluster dynamical status}
As introduced in section~\ref{sec:intro}, the offset between the brightest cluster galaxy and the X-ray peak provides a sensitive probe of the dynamical state of galaxy clusters, particularly at high redshift where ongoing assembly is expected. Our BCG--X-ray peak offset analysis (section~\ref{subsec:peakoffset}) shows that only one out of ten clusters in the high-redshift sample satisfies the criterion for a dynamically ``relaxed'' system ($D_{\mathrm{XP}} < 0.02R_{500}$). The resulting relaxed fraction of $10\%$ (with an upper limit of $\sim 30\%$; table~\ref{tbl:bcg_offset}) is compared here with previous findings.

Although the relaxed fraction is marginally lower than that of low-redshift CAMIRA clusters, the difference is not statistically significant. A similarly low relaxed fraction ($2^{+37}_{-2}\%$) has been reported for high-richness CAMIRA clusters detected in the eFEDS field by eROSITA \citep{Ota23}. These results indicate that high-$z$ optically selected clusters commonly exhibit disturbed morphologies.

A consistent trend has also been reported for X-ray--selected distant clusters. Using the XMM-Newton Distant Cluster Project (XDCP) sample, \citet{Fassbender11} found that BCG--X-ray centroid offsets at $z>0.9$ are systematically larger than those in local clusters, indicating that many high-redshift systems are not yet dynamically relaxed. In contrast, based on a \textcolor{blue}{SZ-selected} sample of the most massive clusters identified by the South Pole Telescope (SPT), \citet{McDonald17} reported little redshift evolution in X-ray morphology out to $z\sim1.9$, suggesting that the most massive clusters can establish regular intracluster medium structures at early cosmic times.

For X-ray--selected clusters in the eFEDS region, \citet{Ghirardini22} introduced a relaxed-score metric and found that the relaxed fraction decreases from $\sim 50\%$ at low redshift to $\sim 30$--$35\%$ at higher redshifts. Our results are consistent with this trend and further support the picture outlined in section~1, in which optically selected high-redshift clusters tend to be dynamically younger than the most massive systems identified in X-ray-- or SZ-selected samples. Taken together, these comparisons imply that high-redshift clusters are dynamically young, likely reflecting the higher merger rate in the early Universe.

\subsection{Redshift evolution of scaling relations}
In section~\ref{subsec:scaling}, we examined the redshift dependence of the $L-T$ and $L-M$ scaling relations using hierarchical Bayesian regression. Although allowing redshift evolution (either free or self-similar) results in slightly lower AIC values, the improvements over the no-evolution model are marginal, and the evolution parameter $c$ remains poorly constrained. Across all models, the best-fit slopes and normalizations are statistically consistent with one another.

These findings indicate that the data do not require significant deviations from self-similar evolution \citep{Kaiser86}. The derived slopes agree with previous analyses of both low-$z$ CAMIRA clusters \citep{Ota20} and high-richness eROSITA-selected clusters \citep{Ota23}, which found $L-T$ and $L-M$ slopes of $2.08 \pm 0.46$ and $1.52 \pm 0.34$, respectively.

Our $N-M$ relation is also consistent with earlier determinations, including $M \propto N^{1.4}$ from WL calibration \citep{Okabe19} and $N \propto M^{0.92}(1+z)^{-0.48}$ from a multi-wavelength analysis of clusters over a wide redshift range \citep{Chiu20}. These comparisons demonstrate the robustness of richness-based mass proxies up to $z \sim 1$.

Recent stacking analyses of $\sim 1000$ CAMIRA clusters in the eFEDS field \citep{Nguyen25} also yield $L-M$ and $L-N$ slopes consistent with ours, further confirming the stability of the derived relations across cluster mass and selection methods.

Theoretical predictions based on the revised baseline model of \citet{Fujita19}, which incorporates the mass-concentration relation, suggest shallower slopes ($L \propto T^{1.6\text{--}1.8}$ and $L \propto M^{1.1\text{--}1.2}$) and weaker redshift dependence than the classical self-similar model. Our results for the no-evolution case ($c=0$) are broadly consistent with these predictions within statistical uncertainties.

Compared with the heterogeneous X-ray cluster compilation of \citet{Reichert11}, our slopes and (weak) redshift dependence are consistent within errors. Their steeper $L-T$ slope ($2.70 \pm 0.24$) and negative evolution parameter contrast with our shallower slope ($2.09 \pm 0.51$) and poorly constrained positive $c$, but the uncertainties overlap. These results further confirm that current observational data provide only limited sensitivity to mild redshift evolution. A more definitive test will require the homogeneous and statistically powerful samples expected from future all-sky eROSITA releases (eRASS).

We note that the relatively small sample size reflects both the observational difficulty of obtaining deep X-ray data and the intrinsic rarity of massive clusters at $z \sim 1$, and our conclusions are therefore intentionally framed in terms of trends rather than definitive measurements.

\subsection{AGN fraction in clusters}\label{subsec:agn_fraction}

\begin{figure*}[htb]
\begin{center}
\includegraphics[width=\textwidth]{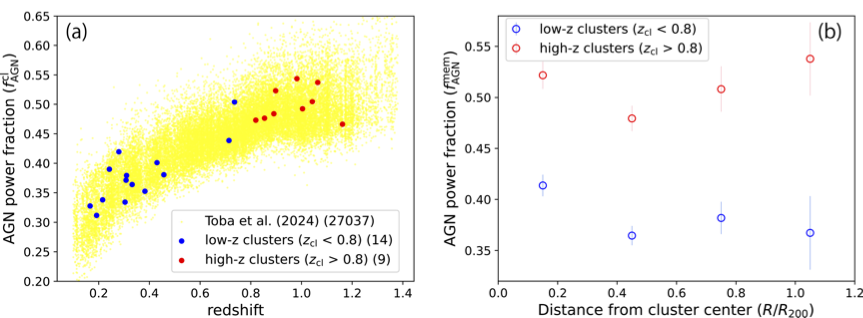}
\end{center}
\caption{(a) Cluster-averaged AGN power fraction ($f^{\mathrm{cl}}_{\mathrm{AGN}}$) as a function of redshift for the sample of 23 galaxy clusters cross-matched with the CAMIRA s21a catalog \citep{Toba24}. Blue and red symbols correspond to low-redshift ($z_{\mathrm{cl}} < 0.8$) and high-redshift ($z_{\mathrm{cl}} > 0.8$) clusters, respectively. (b) Radial profile of the AGN power fraction ($f^{\mathrm{mem}}_{\mathrm{AGN}}$) of member galaxies as a function of normalized cluster-centric radius ($R/R_{200}$). The blue and red crosses represent the average profiles for low-$z$ and high-$z$ clusters, respectively. 
{Alt text: Two panel figure showing active galactic nucleus power fractions in galaxy clusters. Panel a displays a scatter plot of cluster averaged power fraction as a function of redshift for two groups of clusters, low redshift and high redshift, plotted with different symbol types. Panel b shows the radial profile of the member galaxy power fraction as a function of the normalized cluster centric radius, again with two symbol sets representing the two redshift groups. }}\label{fig:agn_fraction}
\end{figure*}

We also investigated the AGN fraction among cluster member galaxies and its dependence on redshift and cluster-centric radius. Using AGN classifications from the CAMIRA s21a catalog \citep{Toba24}, we find that high-$z$ clusters ($z>0.8$) exhibit systematically higher AGN fractions than their low-$z$ counterparts. The enhancement is strongest in the outskirts ($R/R_{200} > 0.5$), where AGN activity rises with radius (figure~\ref{fig:agn_fraction}).

This radial behavior suggests that AGN activity is particularly prevalent during the early stages of cluster assembly, especially in infalling regions where galaxies may retain more cold gas and experience interactions such as ram-pressure compression or minor mergers. These conditions facilitate black hole accretion and trigger AGN activity. The elevated AGN incidence at high redshift is also consistent with theoretical expectations, as merger rates and gas fractions are both higher at earlier cosmic times.

A similar trend has been reported for X-ray-selected distant clusters. Using the XDCP sample, \citet{Fassbender11} showed that clusters at $z>0.9$ host a significantly higher fraction of cluster-associated AGN—particularly luminous radio sources—by a factor of $\sim$2.5--5 compared to local systems, supporting the view that enhanced AGN activity is a common feature of dynamically young clusters.

Although the present analysis is limited to AGN demographics and their spatial distribution and does not directly measure the thermal impact of AGN feedback on the ICM, the enhanced AGN fraction and its spatial distribution provide indirect evidence that AGN may influence the thermodynamic state of dynamically young clusters. Hydrodynamical simulations \citep{Truong18} show that different feedback models (stellar, AGN, or non-radiative) produce distinguishable $L_X$--$T$ relations at high redshift. In particular, the variation in normalization at $z=2$ can reach $\sim 40\%$ (see \cite{Zhang20} figure~13), potentially detectable with the temperature precision expected from Athena's Wide Field Imager \citep{Cruise25}.

Given the elevated AGN fraction in our high-$z$ sample, expanding the analysis to lower-mass clusters at similar redshifts will enable a more complete census of AGN demographics. We therefore plan a systematic X-ray study of CAMIRA clusters across a wide range of richness and redshift using data from the eROSITA All-Sky Survey (eRASS; \cite{Merloni24}). This will allow us to test whether AGN activity correlates with variations in the normalization or intrinsic scatter of ICM scaling relations, particularly in low-mass or dynamically young systems, and thereby clarify the role of AGN feedback in the thermal evolution of galaxy clusters.

\section{Summary}
We presented results from deep \textit{XMM-Newton} observations of ten high-redshift ($0.81 < z < 1.17$) galaxy clusters selected from the CAMIRA catalog for their high richness ($\hat{N}_{\mathrm{mem}} > 40$). These massive systems ($M_{500} \gtrsim 3 \times 10^{14}\, h^{-1} M_\odot$) provide a well-defined sample for studying the thermodynamic and dynamical properties of the intracluster medium (ICM) at $z \sim 1$.

We performed uniform X-ray imaging and spectral analyses to measure the temperature and bolometric luminosity for each cluster, and quantified their dynamical states using offsets between the BCG and X-ray peak. When available, weak-lensing masses from HSC-Y3 were incorporated to calibrate the scaling relations.

Our main findings are summarized as follows:

\begin{itemize}
  \item ICM detection and morphology: Extended X-ray emission was detected from all ten clusters.  Only one system meets the criterion for a dynamically ``relaxed'' cluster ($D_\mathrm{XP} < 0.02\,R_{500}$), yielding a relaxed fraction of $\sim 10\%$  (upper limit $\sim 30\%$).  This indicates that high-redshift CAMIRA clusters are predominantly dynamically disturbed, consistent with previous optical and X-ray studies.

  \item Scaling relations: Combining the high-$z$ sample with 17 low-$z$ CAMIRA clusters \citep{Ota20} allowed us to derive $N$--$T$, $L$--$T$, $N$--$M$, and $L$--$M$ relations over 
  $0.14 < z < 1.17$.  All relations are consistent with self-similar predictions within uncertainties. Although models permitting redshift evolution show slightly improved AIC values, the improvement is not statistically significant, indicating no strong evidence for redshift-dependent deviations in the scaling relations.

  \item AGN activity: High-redshift clusters exhibit a significantly elevated AGN fraction, particularly in their outskirts ($R > 0.5\,R_{200}$), compared to low-$z$ systems. This suggests enhanced AGN triggering during early cluster assembly, which may be related to the ICM thermodynamic state of dynamically young systems.
\end{itemize}

Overall, our results indicate that massive clusters at $z \sim 1$ already follow mature ICM scaling relations, despite exhibiting highly disturbed morphologies and elevated AGN activity. Deep X-ray observations such as those presented here are essential for understanding the thermodynamic evolution of high-redshift clusters and provide a valuable benchmark for interpreting the rapidly growing population of clusters detected by eROSITA.

\bibliographystyle{pasj}
\bibliography{ref}

\begin{ack}
The Hyper Suprime-Cam (HSC) collaboration includes the astronomical communities of Japan and Taiwan, and Princeton University.  The HSC instrumentation and software were developed by the National Astronomical Observatory of Japan (NAOJ), the Kavli Institute for the Physics and Mathematics of the Universe (Kavli IPMU), the University of Tokyo, the High Energy Accelerator Research Organization (KEK), the Academia Sinica Institute for Astronomy and Astrophysics in Taiwan (ASIAA), and Princeton University.  Funding was contributed by the FIRST program from the Japanese Cabinet Office, the Ministry of Education, Culture, Sports, Science and Technology (MEXT), the Japan Society for the Promotion of Science (JSPS), Japan Science and Technology Agency (JST), the Toray Science Foundation, NAOJ, Kavli IPMU, KEK, ASIAA, and Princeton University.

This work was supported in part by the Fund for the Promotion of Joint International Research, JSPS KAKENHI Grant Number 20K04027, 23H00121(NO), JP25H00662, JP24K00684(MO). 
NO acknowledges partial support by the Organization for the Promotion of Gender Equality at Nara Women's University.
\end{ack}

\end{document}